\documentclass[preprintnumbers,showpacs,amsmath,amssymb,floatfix,9pt,prd,onecolumn,
superscriptaddress,nofootinbib]{revtex4}
\usepackage{CJK}
\usepackage{graphicx}
\usepackage{latexsym}
\usepackage{epsfig}
\usepackage{enumerate}
\usepackage{amssymb}
\usepackage[utf8]{inputenc}
\usepackage{xcolor}
\usepackage[%
  colorlinks=true,
  urlcolor=blue,
  linkcolor=red,
  citecolor=blue
]{hyperref}
\usepackage{orcidlink}

\begin{document}
\title{Noncommutative black hole in de Rham-Gabadadze-Tolley like massive gravity }

\author{Piyali Bhar \orcidlink{0000-0001-9747-1009}}
\email[Email: ]{piyalibhar90@gmail.com}
\affiliation{Department of
Mathematics, Government General Degree College Singur, Hooghly, West Bengal 712 409,
India}

\author{Dhruba Jyoti Gogoi\orcidlink{0000-0002-4776-8506}}
\email[Email: ]{moloydhruba@yahoo.in}

\affiliation{Department of Physics, Moran College, Moranhat, Charaideo 785670, Assam, India}
\affiliation{Theoretical Physics Division, Centre for Atmospheric Studies, Dibrugarh University, Dibrugarh}

\author{Supakchai Ponglertsakul}
\email[Email: ]{supakchai.p@gmail.com}
\affiliation{Strong Gravity Group, Department of Physics, Faculty of Science, Silpakorn University, Nakhon Pathom 73000, Thailand}

\begin{abstract}
   We examine the behavior of non-commutative Schwarzschild black holes in the context of massive gravity. According to the investigation, corresponding to a minimal mass, the black hole can have two horizons, one horizon, or no horizon at all. Our results imply the existence of a stable black hole remnant, whose mass can be uniquely calculated in terms of the non-commutative parameter $\theta$ and the graviton mass m. Thermodynamic features such as heat capacity and Hawking temperature are studied. We also examine a scalar linear perturbation on the black hole. Quasinormal frequencies are computed via Wentzel–Kramers–Brillouin (WKB) method with Pad\'e improvement. All quasinormal frequencies considered in this work have a negative imaginary part. In the eikonal limit, we investigate the angular velocity and the Lyapunov exponent as a function of $M/\sqrt{\theta}$. 
   Additionally, we explore the black hole's shadow across various model parameters. Our findings indicate that non-commutativity leads to a reduction in the black hole's shadow, with this effect exhibiting a nonlinear relationship. Furthermore, we observe that the inclusion of a massive graviton in the theory results in an increase in the black hole's shadow radius, particularly at greater observer distances. 
\end{abstract}
\maketitle

\section{Introduction}

The phenomenology of the dRGT massive gravity model is very fascinating. The dRGT black hole (BH) and black string (BS) solutions differ from the standard general relativity solution in the presence of the heavy graviton because the graviton mass plays a role in this situation. Similar terms can also be found in Bergshoeff, Hohm, and Townsend (BHT) massive gravity \cite{Oliva:2009ip} and the minimal theory of massive gravity \cite{DeFelice:2018vza}. 
Recently black hole (BH), black string (BS), and, more recently, rotating black string solutions have been found and examined in the context of the dRGT massive gravity.

For a generic choice of the parameters of dRGT massive gravity, Ghosh et al. \cite{Ghosh:2015cva} provided a precise spherical black hole solution and also discussed the black hole's thermodynamics and phase structure (for the charged case) in the grand canonical and canonical ensembles. Tannukij et al. \cite{Tannukij:2017jtn} proposed a cylindrically symmetric black string solution (both charged and uncharged) to the nonlinear ghost-free massive gravity discovered by de Rham, Gabadadze, and Tolley (dRGT). According to the author, this ``dRGT black string" is considered as a generalization of Lemos's black string solution. Furthermore, since the graviton mass contributes to both the global monopole term and the cosmological-constant term, the dRGT black string solution encompasses additional classes of black string solutions, such as the monopole-black string ones. A rotating black string solution in de Rham-Gabadadze-Tolley (dRGT) massive gravity is discussed in \cite{Ghosh:2019eoo}. This solution has two more variables that characterize the structure of graviton mass, making it a kind of generalized variant of the rotating anti-de Sitter (AdS)/dS black string solution. Refs. \cite{Burikham:2017gdm,Ponglertsakul:2018smo,Boonserm:2017qcq, Nieuwenhuizen:2011sq, Boonserm:2019mon,Panpanich:2019mll,Wuthicharn:2019olp} contain some other intriguing works on black holes under massive gravity.


Since non-commutative spacetime in gravity theories can be considered as an alternative to quantum gravity, it has gained attention as an active research topic in recent years \cite{Nicolini:2008aj, Snyder:1946qz}. Several studies have been carried out about non-commutative gravity. Specifically, the impact of non-commutativity on black hole physics has drawn significant interest, mainly due to the increased abundance of the final stage of the noncommutative black hole. It is familiar that the non-commutativity can be applied to General Relativity by changing the matter source \cite{Nicolini:2005vd}, eliminating pointlike structures in favour of smeared objects in flat spacetime \cite{Smailagic:2003yb, Smailagic:2003rp}. Consequently, mass density is changed to introduce non-commutativity by substituting a Gaussian distribution for the Dirac delta function \cite{Nicolini:2005vd}. The Schwarzschild black hole, inspired by four-dimensional non-commutative geometry, was first postulated by Nicolini \cite{Nicolini:2005vd}. This black hole solution was then extended to include the scenario of charged \cite{Ansoldi:2006vg}. In 2010, this study was further extended to the general case of charged rotating non-commutative black holes by Modesto and Nicolini \cite{Modesto:2010rv}. Hamil et al. \cite{Hamil:2024ppj} investigated the shadow images in the presence of plasma and used Hawking temperature, entropy, and specific heat functions to study the thermodynamics of non-commutative Schwarzschild black holes embedded in quintessence matter. They also explored phase transition and stability features and the quasinormal modes in WKB and Mashhoon approximations. Finally, they thoroughly discussed the impacts of quintessence matter and non-commutative parameter on the black hole spacetime. Heidari et al. \cite{Heidari:2023bww} computed the thermodynamic parameters of the system and examined the gravitational signatures of a non-commutative stable black hole, comparing their findings with those found in recent research. They used the WKB approximation and the Pöschl–Teller fitting method to investigate the quasinormal modes of massless scalar perturbations. Additionally, they looked at the geodesics of massive and massless particles, showing how the non-commutative parameter $\theta$ greatly affects event horizons and light pathways. Additionally, they computed the shadows, demonstrating that larger values of $\theta$ correspond to larger shadow radii. Yan and colleagues \cite{Yan:2023pxj} computed the quasinormal modes (QNMs) of a charged non-commutative black hole using three different methods: scalar, electromagnetic, and gravitational fields. They discussed the effects of charge $Q$ and the non-commutative parameter $\theta$ on QNMs in various domains. After that, scientists used the event horizon telescope's constraints on the shadow radii of M87* and Sgr A* to compute the black hole's shadow radius and provided the valid range of $\theta$ and Q. Saleem and Aslam \cite{Saleem:2023pyx} examined the observable behavior and shadow of the non-commutative (NC) charged Kiselev black hole, which was surrounded by various profiles of accretions. The authors selected particular values for the model parameters to produce the BH shadow profile, and they concluded that changes in each parameter directly affected the BH size and light trajectories. In ref \cite{Iftikhar:2023wug}, Iftikhar investigated the shadow cast by a non-commutative rotating Hayward black hole. Based on the study, the author demonstrates that the spin, non-commutative parameter, and parameter $g$ of the mentioned black hole determine both the apparent shape and size of the shadow. Both the size of the shadow and the non-commutative parameter decrease with $g$. Additionally, the author discovered that at large values of $g$ and spin, the shape of the shadow deviates from a complete circle. Zeng et al.\cite{Zeng:2022fdm} examined the light rings and observational characteristics of the BH shadow encircled by various accretion flow models in their system of consideration. They next looked at how model factors affected the space-time structure and observational display of BHs. Investigation on wormholes under non-commutative geometry were studied in \cite{Tayde:2023gub,Farooq:2023rsp,Pradhan:2023vhn,Chalavadi:2023zcw,Debnath:2023yfm,Tayde:2023xbm, Kavya:2023mwv,Rahaman:2012pg,Bhar:2014ooa}


The evolution of a linear matter wave around black hole is generally governed by quasi-normal modes (QNMs). QNMs are associated to complex frequency (quasi-normal frequency) where real part describes emission frequency and imaginary part represents the decaying time scale. Remarkably, quasinormal frequency can be uniquely determined by black hole's mass and angular momentum which allows one to obtain various properties of black holes by studying QNMs of black holes. The QNMs study can be traced back to 1970 where the first QNMs of Schwarzschild black hole was investigated \cite{Vishveshwara:1970zz}. Since then, an astronomical amount of studies devoted to black hole's QNMs have been explored (see for example, \cite{Press:1971wr,Kokkotas:1999bd,Konoplya:2011qq, Gogoi:2023lvw, Gogoi:2023ffh, Konoplya:2002wt, Konoplya:2003ii,  Konoplya:2017wot, Konoplya:2019hlu, Sekhmani:2023ict, Gogoi:2023fow, Gogoi:2024vcx, Okyay:2021nnh, Pantig:2022gih, Pedrotti:2024znu}). In addition, charged black holes in the dRGT massive gravity are superradiantly unstable \cite{Burikham:2017gdm}. In contrast, no instability modes are found in the case of neutral black string in the dRGT massive gravity \cite{Ponglertsakul:2018smo}. On the other hand, within noncommutative framework, various QNMs of black holes are explored e.g. the Schwarzschild black hole \cite{Giri:2006rc,Liang:2018uyk,Liang:2018nmr,Campos:2021sff,Zhao:2023uam}, the Reissner-Nordstr\"om black hole \cite{Ciric:2017rnf} and the BTZ black hole \cite{Gupta:2017lwk,Gupta:2015uga}.

The study of black holes reveals their event horizon as a crucial boundary beyond which no particles, even light, can escape due to intense gravitational forces \cite{Synge:1966okc}. Matter falling into a black hole, known as accretion, spirals inward, emitting radiation across various frequencies, including detectable radio waves, forming a bright background with a dark region called the black hole shadow \cite{Luminet:1979nyg}. Pioneered by Falcke et al. \cite{Falcke:1999pj}, recent advancements, notably by the Event Horizon Telescope, have allowed the imaging of black hole shadows in galaxies like Messier 87* (M 87*) and Sagittarius A* (Sgr. A*) \cite{EventHorizonTelescope:2019dse,EventHorizonTelescope:2022xnr}. These observations offer insights into photon sphere distortion caused by the non-spherical horizons of rotating black holes \cite{Ovgun:2018tua, EventHorizonTelescope:2021dqv, Belhaj:2020okh,Belhaj:2020rdb, Belhaj:2022kek, gogoi_joulethomson_2023, Ovgun:2020gjz,Ovgun:2019jdo,Kuang:2022xjp,Kumaran:2022soh,Mustafa:2022xod,Cimdiker:2021cpz,Okyay:2021nnh,Atamurotov:2022knb,Pantig:2022qak,Abdikamalov:2019ztb,Abdujabbarov:2016efm,Atamurotov:2015nra,Papnoi:2014aaa,Abdujabbarov:2012bn,Atamurotov:2013sca,Cunha:2018acu,Gralla:2019xty}. These advancements have led to a comprehensive study of shadows in various modified gravities, including comparisons with M87* and Sgr. A* observations, providing valuable geometric data \cite{Pantig:2022gih,Lobos:2022jsz, gogoi_joulethomson_2023, Uniyal:2023inx,Panotopoulos:2021tkk,Panotopoulos:2022bky,Khodadi:2022pqh,Khodadi:2021gbc,Zhao:2023uam,Khodadi:2020jij,Khodadi:2020gns,Vagnozzi:2022moj,Khodadi:2022ulo}. Furthermore, the study of black hole shadows has revealed insights into thermodynamic phase transitions, with the shadow radius serving as a key tool for studying thermodynamic black hole systems \cite{Zhang:2019glo}. Additionally, a significant relationship has been established between black hole shadows and quasi-normal modes, particularly in the eikonal limit, demonstrating that the real parts of quasi-normal modes are connected to the shadow radii of black holes \cite{Jusufi:2019ltj,Jusufi:2020dhz,Jusufi:2020mmy}.

In this work, we investigate thermodynamics properties, perturbation and optical properties of noncommutative black hole in dRGT massive gravity. Hence, this paper is organized as follows. Section~\ref{sec:action}, we discuss the action and modified field equation within the framework of the dRGT massive gravity. Noncommutative black hole is introduced in Section~\ref{sec:model}. Then, horizon, asymptotic structure and curvature scalars of the black hole solution are discussed. Later, thermodynamics properties such as temperature and heat capacity are explored. In Section~\ref{sec:QNMs}, we calculate quasinormal frequencies of noncommutative black holes and investigate the eikonal limit. Optical appearance i.e., black hole's shadow is studied in Section~\ref{sec:shadow}. Finally, we summarize the results and discuss the novel finding in the conclusions section.

\section{action and field equation in massive gravity}\label{sec:action}
To examine the black hole model within the framework of de Ramadanze–Tolley (dRGT) massive gravity, we take into consideration the following action:
\begin{equation}
\mathcal{I}=\frac{1}{16\pi G}\int d^{4}x\sqrt{-g}\left[ \mathcal{R}
+m^{2}\sum_{i}^{4}c_{i}\mathcal{U}_{i}(g,f)\right] +L_{\text{m}},
\label{Action}
\end{equation}%
where $L_{\text{m}}$ is the matter Lagrangian, $g$ is the determinant of the metric tensor $g_{\mu\nu}$. $f$ and $g$ are fixed symmetric tensors and metric tensors, respectively, and $\mathcal{U}_{i}$ is the self-interacting potential of the graviton with mass $m$, $c_{i}$'s are constant. The potential $\mathcal{U}_{i}$ in this case is defined as,
\begin{eqnarray}
\mathcal{U}_{1} &=&\left[ \mathcal{K}\right] ,\;\;\;\;\;\;\;\mathcal{U}_{2}=%
\left[ \mathcal{K}\right] ^{2}-\left[ \mathcal{K}^{2}\right] ,  \nonumber \\
\mathcal{U}_{3} &=&\left[ \mathcal{K}\right] ^{3}-3\left[ \mathcal{K}\right] %
\left[ \mathcal{K}^{2}\right] +2\left[ \mathcal{K}^{3}\right] ,  \nonumber \\
\mathcal{U}_{4} &=&\left[ \mathcal{K}\right] ^{4}-6\left[ \mathcal{K}^{2}%
\right] \left[ \mathcal{K}\right] ^{2}+8\left[ \mathcal{K}^{3}\right] \left[
\mathcal{K}\right] +3\left[ \mathcal{K}^{2}\right] ^{2}-6\left[ \mathcal{K}%
^{4}\right] .  \nonumber
\end{eqnarray}
$\left[\mathcal{K}\right]$ represents the trace of the metric $\mathcal{K}^{\mu}_{\nu} = \delta^{\mu}_{\nu} - \sqrt{g^{\mu\sigma}f_{ab}\partial_{\sigma}\phi^a\partial_{\nu}\phi^b}$. The $f_{ab}$ is fiducial or reference metric to be defined later. The four Stuckelberg fields take the form $\phi^a=x^\mu \delta^{a}_\mu$. 

By varying the action \eqref{Action} for the metric tensor $g^{\mu}_{\nu}$, the equation of motion for this gravity can be obtained as follows:
\begin{equation}
R_{\mu \nu }-\frac{1}{2}R g_{\mu \nu }+m^{2}\chi _{\mu \nu}=\frac{8\pi G}{c^4}T_{\mu \nu },  \label{Field equation}
\end{equation}%
the expression for $\chi _{\mu \nu } $ is represented by,
\begin{eqnarray}
\chi _{\mu \nu } &=&-\frac{c_{1}}{2}\left( \mathcal{U}_{1}g_{\mu \nu }-%
\mathcal{K}_{\mu \nu }\right) -\frac{c_{2}}{2}\left( \mathcal{U}_{2}g_{\mu
\nu }-2\mathcal{U}_{1}\mathcal{K}_{\mu \nu }+2\mathcal{K}_{\mu \nu
}^{2}\right)  \nonumber \\
&&  \nonumber \\
&&-\frac{c_{3}}{2}(\mathcal{U}_{3}g_{\mu \nu }-3\mathcal{U}_{2}\mathcal{K}%
_{\mu \nu }+6\mathcal{U}_{1}\mathcal{K}_{\mu \nu }^{2}-6\mathcal{K}_{\mu \nu
}^{3})  \nonumber \\
&&  \nonumber \\
&&-\frac{c_{4}}{2}\left( \mathcal{U}_{4}g_{\mu \nu }-4\mathcal{U}_{3}%
\mathcal{K}_{\mu \nu }+12\mathcal{U}_{2}\mathcal{K}_{\mu \nu }^{2}-24%
\mathcal{U}_{1}\mathcal{K}_{\mu \nu }^{3}+24\mathcal{K}_{\mu \nu
}^{4}\right),
\end{eqnarray}
and $T_{\mu \nu }$ represents the energy momentum tensor.\\
In ($3 + 1$) dimensions, let us express the line element that describes the interior spacetime of a spherically symmetric distribution of matter as follows:
\begin{equation}\label{line}
ds^{2}=-h(r)dt^{2}+[h(r)]^{-1}dr^{2}+r^{2}(d\Theta^2+\sin^2\Theta d\phi^2),
\end{equation}
where $h(r)=e^{2\phi(r)}$ and $[h(r)]^{-1}=e^{2\lambda(r)}$, and $\phi$ and $\lambda$ depend on `r', i.e., purely radial. Generalizing and using the ansatz in  \cite{Cai:2014znn}, we employ the reference metric that follows:
\begin{equation}
f_{\mu \nu }=\text{diag}(0,0,C^{2},C^{2}\sin ^{2} \Theta ),  \label{f11}
\end{equation} $C$ being a positive constant. 
With the reference metric (\ref{f11}), we have \cite{Cai:2014znn},
\begin{equation}
\mathcal{U}_{1}=\frac{2C}{r},\text{ \ \
}\mathcal{U}_{2}=\frac{2C^{2}}{r^{2}},\,\mathcal{U}_{3}=0,\,\mathcal{U}_{4}=0. \nonumber
\end{equation}
We assume that the matter distribution is anisotropic, and therefore we select the most general energy-momentum tensor in the form:
\begin{equation}
T_{\mu}^{ \nu }=(\rho+p_t)u^{\nu}u_{\mu}+p_tg_{\mu}^{\nu}+(p_r-p_t)v^{\nu}v_{\mu},
\label{EMTensorEN}
\end{equation}
The symbols $v^{\nu}$ are the spacelike vector and $u_{\mu}$ represent fluid four velocities. $\rho(r),\, p_r(r)$, and $p_t(r)$ stand for the energy density, radial pressure, and tangential pressure, respectively.
The field equations (assuming $G=c=1$) in de Rham-Gabadadze-Tolley like massive gravity, are given as \cite{Li:2023rkr}:
\begin{eqnarray}
8\pi r^{2}\rho &=&e^{-2\lambda}(2r\lambda'-1)+1+m^2C(c_2C+c_1r),  \label{1} \\
&&  \nonumber \\
8\pi r^{2}p_r &=&e^{-2\lambda}(2r\phi'+1)-1-m^2C(c_2C+c_1r),  \label{2} \\
&&  \nonumber \\
8\pi r p_t &=&e^{-2\lambda}(\phi'+r\phi'^2-\lambda'-r\lambda'\phi'+r\phi'')-\frac{m^2c_1C}{2}.  \label{3}
\end{eqnarray}%
Here the derivative with respect to $r$ is indicated by the prime. The conservation law for the energy–momentum tensor is obtained by using $T_{\mu;\nu}^{\nu}=0$ as,
\begin{eqnarray}
    \frac{dp_r}{dr}=-(\rho+p_r)\frac{d\phi}{dr}+\frac{2}{r}(p_t-p_r)=0.
\end{eqnarray}
The goal of the upcoming section is to derive the black hole model by solving the field equations in the presence of non-commutative geometry.

\section{Black hole model}\label{sec:model}
In ($3+1$) dimensions, when a static, spherically symmetric, noncommutative distributed, particle-like gravitational source is taken into consideration, one finds,
\begin{eqnarray}\label{rho1}
\rho = \frac{M}{(4\pi\theta)^{\frac{3}{2}}}e^{-\frac{r^2}{4\theta}},
\end{eqnarray}
Here, $M$ denotes the source's total mass. The matter distribution in models inspired by noncommutative geometry is smeared on a scale that is directly related to the noncommutativity parameter, $\theta$. The assertion is that the uncertainty principle between coordinates prevents the matter from being fully localized in the presence of noncommutativity. According to phenomenological findings, noncommutativity is not apparent at presently accessible energies, i.e., $\theta<10^{-16}$ cm.

In addition, to obtain a black hole solution, one requires $g_{tt}=(g_{rr})^{-1}$ that implies, $p_r=-\rho$. Using the expression of $\rho$ given in (\ref{rho1}), we solve the equation (\ref{1}) for $e^{-2\lambda}$ which is obtained as,
\begin{eqnarray}\label{elambda}
e^{-2\lambda}&=&1 + C m^2 \left(C c_2 + \frac{c_1 r}{2}\right)+\frac{2M}{\sqrt{\theta \pi}}e^{-\frac{r^2}{4\theta}}-\frac{2M}{r}\text{erf}\left(\frac{r}{2\sqrt{\theta}}\right),
\end{eqnarray}
where $\text{erf}(r)$ is the error function. 
Using field equations (\ref{1})-(\ref{3}), we get,
\begin{eqnarray}
e^{2\phi}=e^{-2\lambda}&=&1 + C m^2 \left(C c_2 + \frac{c_1 r}{2}\right)+\frac{2M}{\sqrt{\theta \pi}}e^{-\frac{r^2}{4\theta}}-\frac{2M}{r}\text{erf}\left(\frac{r}{2\sqrt{\theta}}\right). \label{bhmetric}
\end{eqnarray}
This metric is clearly not an asymptotically flat spacetime. In the absence of graviton mass $m\rightarrow0$, one can readily be obtained as,
\begin{eqnarray}
e^{2\phi}=e^{-2\lambda}&=&1 +\frac{2M}{\sqrt{\theta \pi}}e^{-\frac{r^2}{4\theta}}-\frac{2M}{r}\text{erf}\left(\frac{r}{2\sqrt{\theta}}\right),
\end{eqnarray}
which is the expression of the non-commutative Schwarzschild black hole proposed by Nicolini \cite{Nicolini:2005zi}, and, for
$\frac{r}{\sqrt{\theta}} \rightarrow \infty$, and $m\rightarrow 0$, the standard Schwarzschild metric $e^{-2\lambda}=e^{2\phi}=1-\frac{2M}{r}$ can be constructed from (\ref{elambda}).\\
The radial and transverse pressures can be obtained as,
\begin{eqnarray}
p_r&=&-\frac{M}{(4\pi\theta)^{\frac{3}{2}}}e^{-\frac{r^2}{4\theta}},\\
p_t&=&\frac{M (r^2 - 4 \theta)}{32 \pi^{\frac{3}{2}} \theta^{\frac{5}{2}}}e^{-\frac{r^2}{4\theta}}
\end{eqnarray}
Analyzing the geometrical and thermodynamical behavior of the solution is worthwhile to comprehend the remaining surprise characteristics fully. 

\subsection{Horizons of the black hole}
\begin{figure}[htbp]
    \centering
    \includegraphics[scale=.7]{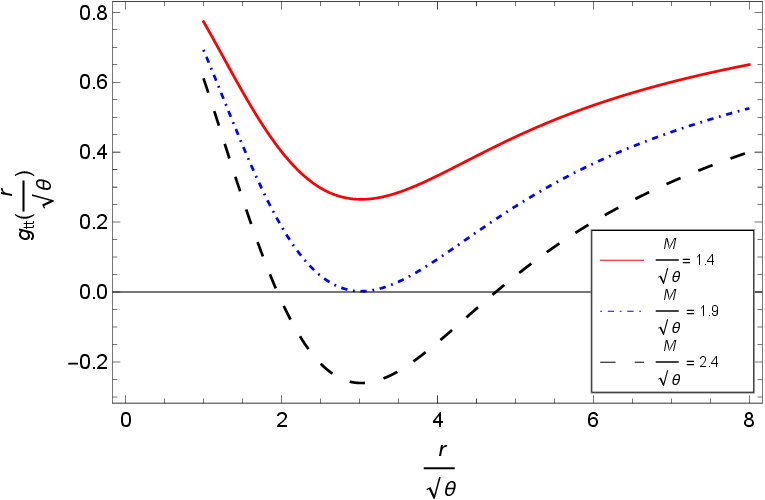}
        \caption{$g_{tt}(\frac{r}{\sqrt{\theta}})$ is shown versus $\frac{r}{\sqrt{\theta}}$. The curves are drawn for various values of $\frac{M}{\sqrt{\theta}}$. For $M=1.4 \sqrt{\theta}$ no horizon exists. For $M=1.9 \sqrt{\theta}$ one degenerate horizon exists at $r=3.05 \sqrt{\theta}$ and for $M=2.4 \sqrt{\theta}$, two horizon exists. For drawing the plots we have taken $C=0.05,\,m^2c_1=0.01,\,m^2c_2=0.05$, and $\theta=0.01$. \label{fig1}}
\end{figure}

Let us begin our investigation with the black hole's horizon equation. The horizon ($r_H$) of the black hole occurs where $g_{rr}(r_H)=0$. That gives,
\begin{eqnarray}\label{rh}
1 + C m^2 \left(C c_2 + \frac{c_1 r_H}{2}\right)+\frac{2M}{\sqrt{\theta \pi}}e^{-\frac{r_H^2}{4\theta}}-\frac{2M}{r_H}\text{erf}\left(\frac{r_H}{2\sqrt{\theta}}\right)=0
\end{eqnarray}
From Eq. (\ref{rh}), we are unable to find a closed-form solution for $r_H$.  Plotting $g_{tt}$, on the other hand, allows one to mathematically determine the presence of horizon(s) and their radius by identifying intersections with the $\frac{r}{\sqrt{\theta}}$-axis. As seen in Fig.~\ref{fig1}, horizons and their radii are present at the locations where $g_{tt}(\frac{r}{\sqrt{\theta}})$ crosses the $\frac{r}{\sqrt{\theta}}$-axis. Noncommutativity introduces new behavior concerning the regular Schwarzschild black hole, as Fig.~\ref{fig1} illustrates. Several possibilities exist, rather than just one event horizon. The point at which the function $g_{tt}(\frac{r}{\sqrt{\theta}})$ touches with the $\frac{r}{\sqrt{\theta}}$-axis indicates the existence of a minimum mass $M_0 = 1.90\sqrt{\theta}$, below which there is no event horizon. In the borderline situation of $M = M_0$, we discover one single degenerate horizon at $r_H \sim 3.05\sqrt{\theta}$, corresponding to the extremal black hole, while the line element exhibits two separate horizons for $M>M_0$. Additionally, the figure also indicates that for $M<M_0$ no black hole exists. Additionally, Fig.~\ref{fig1} shows that as the mass of the black hole increases, so will the distance between the horizons. 

\begin{figure}[htbp]
    \centering
    \includegraphics[scale=.5]{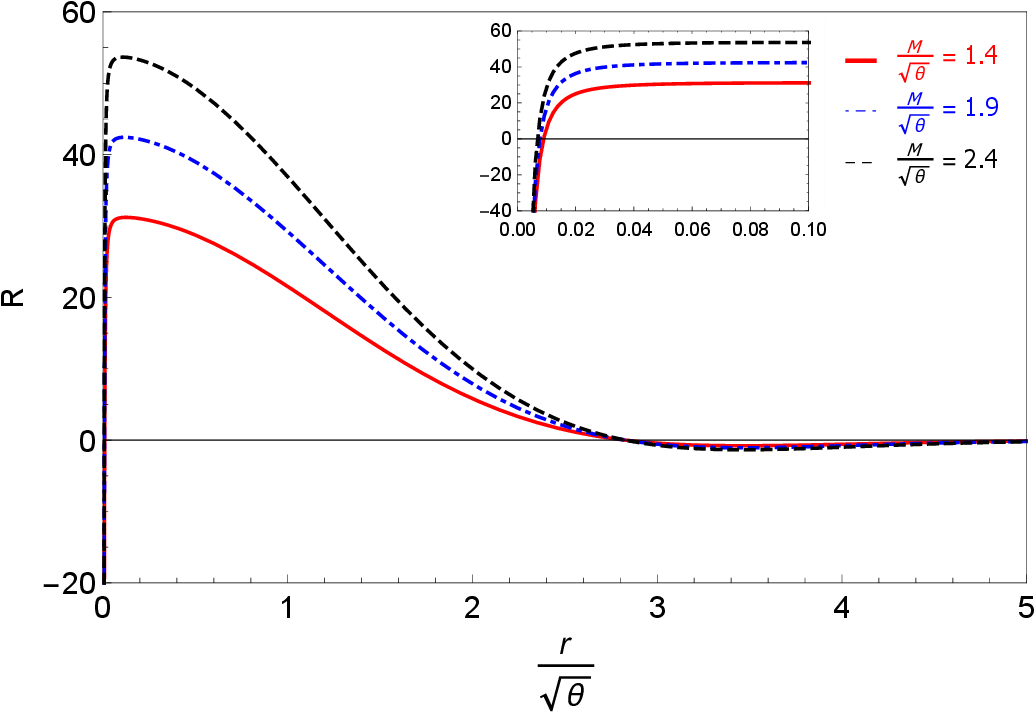}
    \includegraphics[scale=.5]{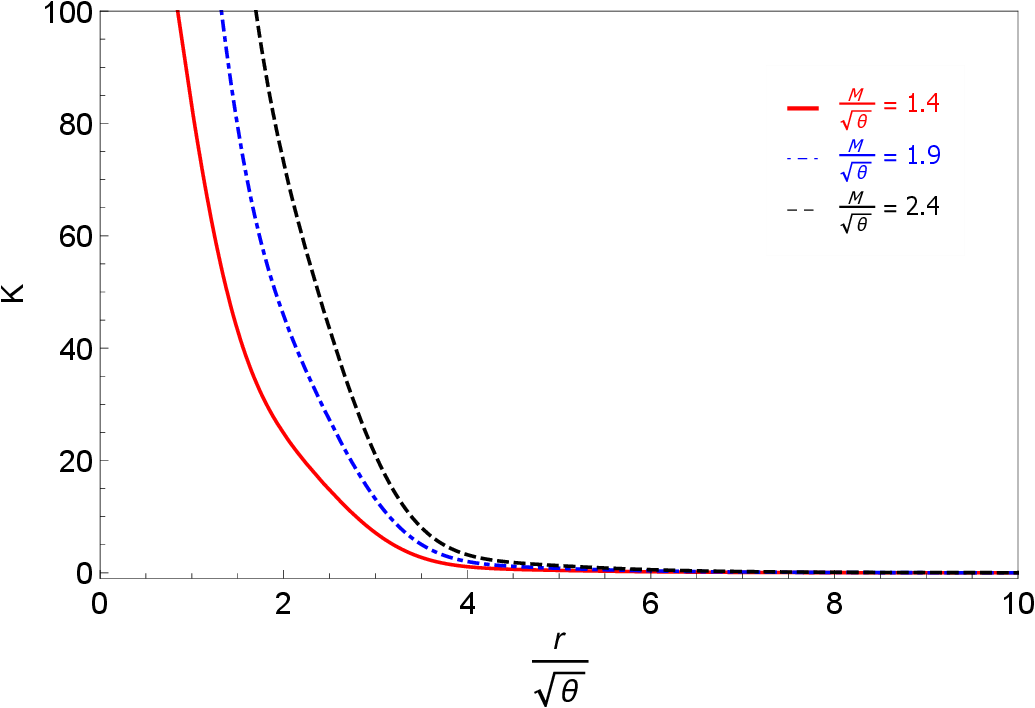}
        \caption{Scalar curvature quantities plot against $\frac{r}{\sqrt{\theta}}$ for $C=0.05,\,m^2c_1=0.01,\,m^2c_2=0.05$, and $\theta=0.01$. Left: the Ricci scalar. Right: the Kretchmann scalar. Note that, a subplot displays behavior of the curvature scalar at small $\frac{r}{\sqrt{\theta}}.$} \label{fig:RandK}
\end{figure}

Intriguingly, the black hole \eqref{rh} is finite as $r\to 0.$ i.e.,
\begin{align}
    e^{2\phi} &\sim \left(1+C^2c_2m^2\right) + O(r),
\end{align}
However, the Ricci scalar $R$ and the Kretchmann scalar $K$ are found to be 
\begin{align}
    R &\sim O(r^{-2}),~~~~~~~~~~~~~~~~~~~ K\sim O(r^{-4}).
\end{align}
as $r\to 0.$ which indicate curvature singularity of the black hole solution. In Fig.~\ref{fig:RandK}, the Ricci scalar and the Kretchmann scalar are displayed. It is apparent that these diverge at $r=0$. Moreover, at the location of event horizon as illustrated in Fig.~\ref{fig1}, these curvature scalars are finite. Therefore despite the black hole solution being regular at $r=0$ there exists curvature singularity. Remarkably, in the $M=1.4\sqrt{\theta}$ case, the solution can be interpreted as a naked singularity. 

\begin{figure}[htbp]
    \centering
    \includegraphics[scale=.5]{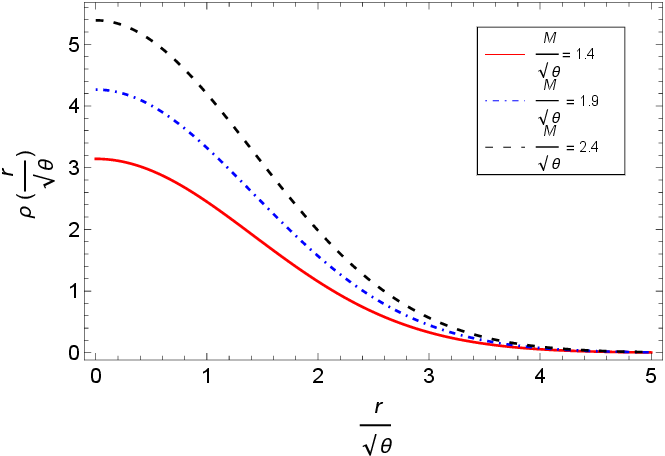}
    \includegraphics[scale=.5]{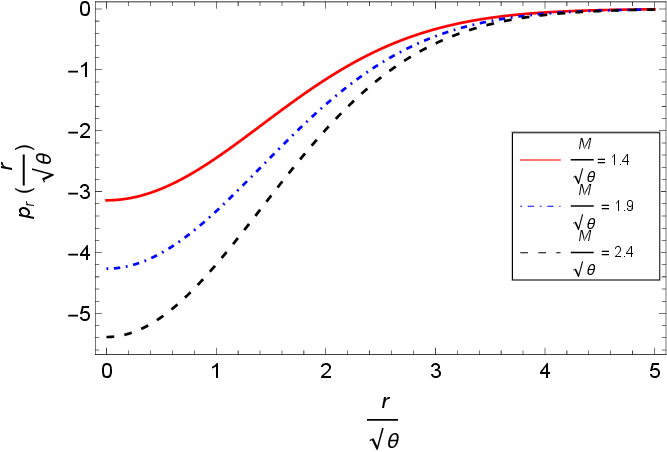}
    \includegraphics[scale=.5]{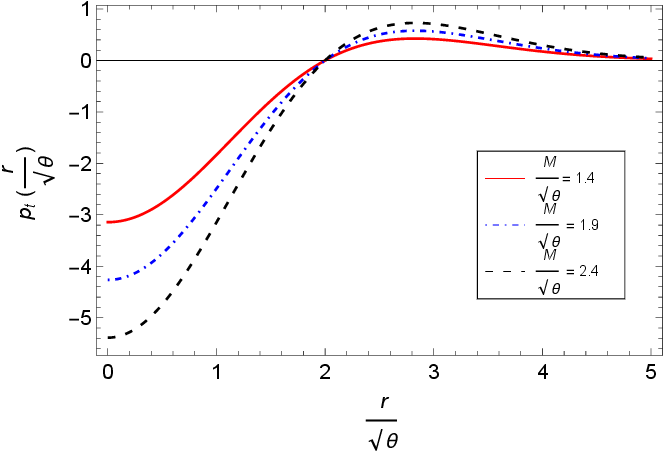}
        \caption{Energy density (left), radial pressure (centre), and transverse pressure (right) plot against $\frac{r}{\sqrt{\theta}}$ for $\theta=0.01$. The other parameters are taken as, $C=0.05,\,m^2c_1=0.01,\,m^2c_2=0.05$, and $\theta=0.01$. }\label{fig:rhopp}
\end{figure}

In addition, the energy density $(\rho)$, radial and transverse pressure ($p_r,p_t$) are plotted as a function of $\frac{r}{\sqrt{\theta}}$ in Fig.~\ref{fig:rhopp}. All of them are finite from within the centre of black hole toward the exterior region. With increasing $\frac{M}{\sqrt{\theta}}$, the energy density becomes larger while both pressures ($p_r,p_t$) are smaller. Moreover as one moves away from the black hole's centre, $\rho$ decreases and $p_r$ increases with $\frac{r}{\sqrt{\theta}}$. The transverse pressure $p_t$ is increasing before reaching its maximum at $r=2.8284\sqrt{\theta}$. A similar type of behavior of the transverse pressure was obtained by Roy Chowdhury et al. \cite{Chowdhury:2019odh} when describing the model of a noncommutative black hole in the Finslerian spacetime and by Rahaman et al \cite{Rahaman:2013gw} when studying BTZ black hole in presence of non-commutative geometry.  It can also be noticed that the strong energy conditions $\rho+p_r+2p_t$ do not hold for $r<2.8284\sqrt{\theta}$ This suggests that the classical description of energy and matter breaks down at the region where quantum phenomena dominate despite nonlinear gravity.

\subsection{Temperature, total mass and the heat capacity of the black hole}
\begin{figure}[htbp]
    \centering
    \includegraphics[scale=.7]{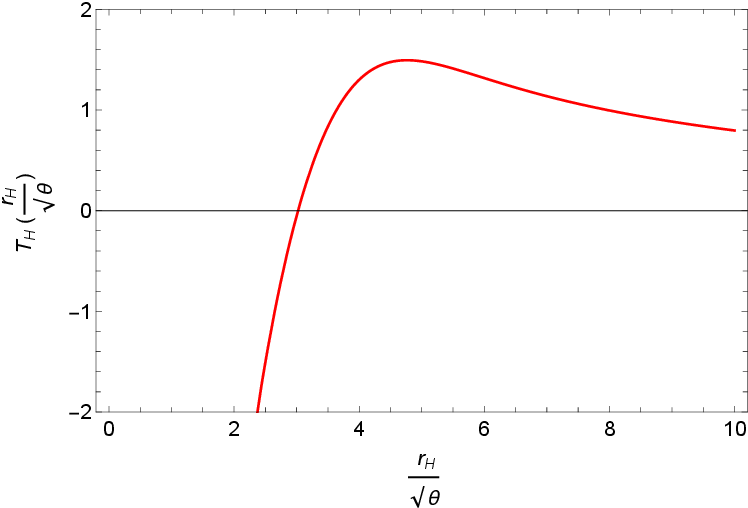}
        \caption{ The profile of Hawking temperature is shown against $\frac{r_H}{\sqrt{\theta}}$. The values of the parameters are taken as, $C=0.05,\,m^2c_1=0.01,\,m^2c_2=0.05$, and $\theta=0.01$.\label{ht}}
\end{figure}

The Hawking temperature of the black hole can be calculated as,
\begin{eqnarray}
    T_H&=&\frac{1}{4\pi}\left(\frac{dg_{tt}}{dr}\right)_{r=r_H},\nonumber\\
    &=&\frac{\frac{r_H^3 \left(C m^2 \left(2 c_2 C+c_1 r_H\right)+2\right)}{\theta  \left(r_H-\sqrt{\pi } \sqrt{\theta } e^{\frac{r_H^2}{4 \theta }} \text{erf}\left(\frac{r_H}{2 \sqrt{\theta }}\right)\right)}+4 C m^2 \left(c_2 C+c_1 r_H\right)+4}{16 \pi  r_H}. \label{temp}
\end{eqnarray}
The plot of the Hawking temperature is shown in Fig.~\ref{ht} for our present noncommutative black hole. The figure indicates a minimum horizon radius $r_0$ since $T_H$ cannot be negative in general, and this radius occurs where $\frac{dg_{tt}}{dr}$ vanishes. One can note that at this radius $r_0$, the Hawking temperature $T_H$ vanishes.

In the absence of massive gravity i.e., $m=0$ and considering the limit $r/\sqrt{\theta}\gg 1$, the expression of $T_H$ $\eqref{temp}$ resemble those of the Schwarzschild black hole 
\begin{align}
    T_H &= \frac{1}{4 \pi  r_H}.
\end{align}

\begin{figure}[htbp]
    \centering
     \includegraphics[scale=.7]{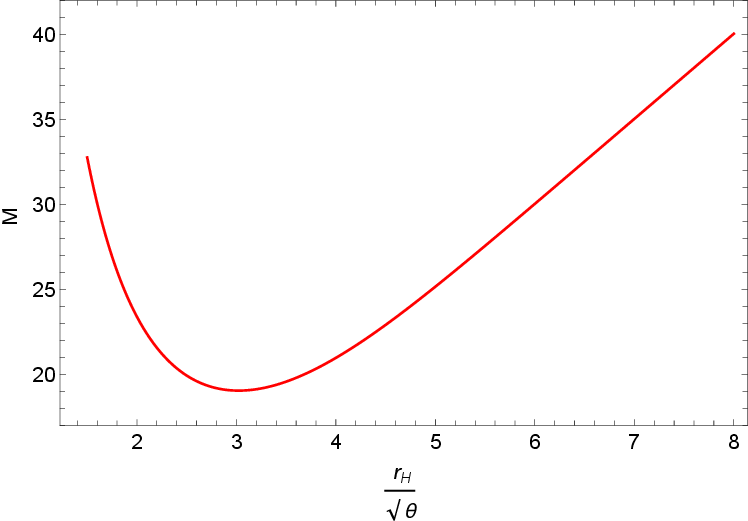}
    \includegraphics[scale=.7]{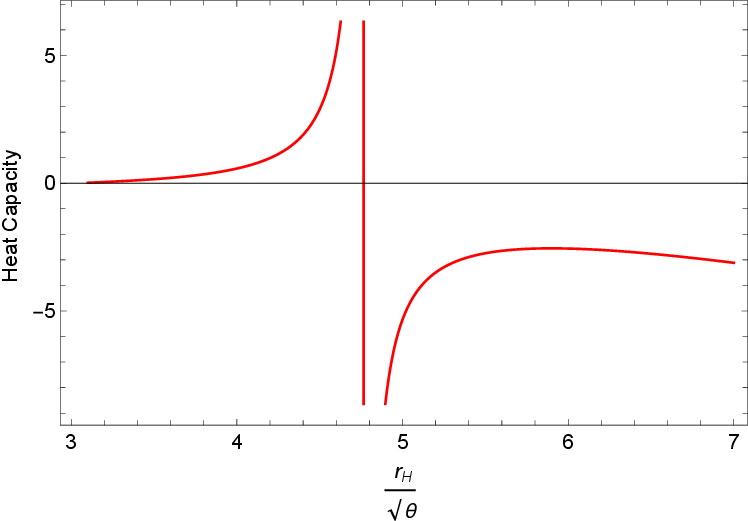}
        \caption{(Left) the total mass $M$ is shown against $\frac{r_H}{\sqrt{\theta}}$, (right) heat capacity is shown against $\frac{r_H}{\sqrt{\theta}}$. For drawing the plots, the values of the parameters are taken as, $C=0.05,\,m^2c_1=0.01,\,m^2c_2=0.05$, and $\theta=0.01$. } \label{totalmass}
\end{figure}

We can obtain the mass in terms of $r_H$ as follows:
\begin{eqnarray}
    M&=&\frac{e^{\frac{r_H^2}{4 \theta}} \sqrt{\pi} r_H (2 + 
   C m^2 (2 C c_2 + c_1 r)) \sqrt{\theta}}{-4 r_H + 4 e^{\frac{r_H^2}{4 \theta}} \sqrt{\pi} \sqrt{\theta}
   erf(\frac{r_H}{2\sqrt{\theta}})}.
\end{eqnarray}
The left panel of Fig.~\ref{totalmass} displays the variation of the overall mass with respect to $\frac{r_H}{\sqrt{\theta}}$. The figure shows that the minimal mass $M_{0}$ validates the degenerate horizon found in Fig.~\ref{fig1}.
The heat capacity of the model can be obtained as
\begin{eqnarray}
    \mathcal{C}=\frac{ \partial M}{\partial r_H}\left(\frac{\partial T_H}{\partial r_H}\right)^{-1},
\end{eqnarray}
We have shown the graphical representation of the heat capacity in Fig~\ref{totalmass} (right panel). The heat capacity is positive for $2.9 \sqrt{\theta}<r_H<4.7\sqrt{\theta}$. Because of their negative heat capacity, black holes are unstable for $r_H>4.7\sqrt{\theta}$. The heat capacity approaches $0$ when $r_H$ approaches $2.9\sqrt{\theta}$.

\section{Quasinormal modes}\label{sec:QNMs}
\begin{figure}[htbp]
    \centering
    \includegraphics[scale=.5]{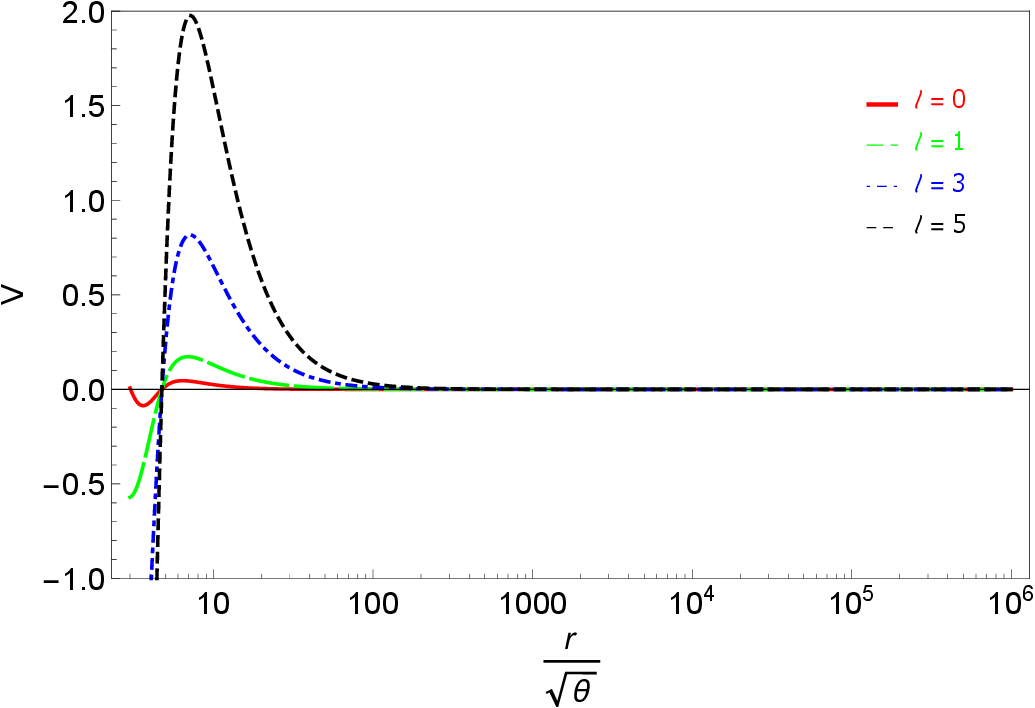}
    \includegraphics[scale=.5]{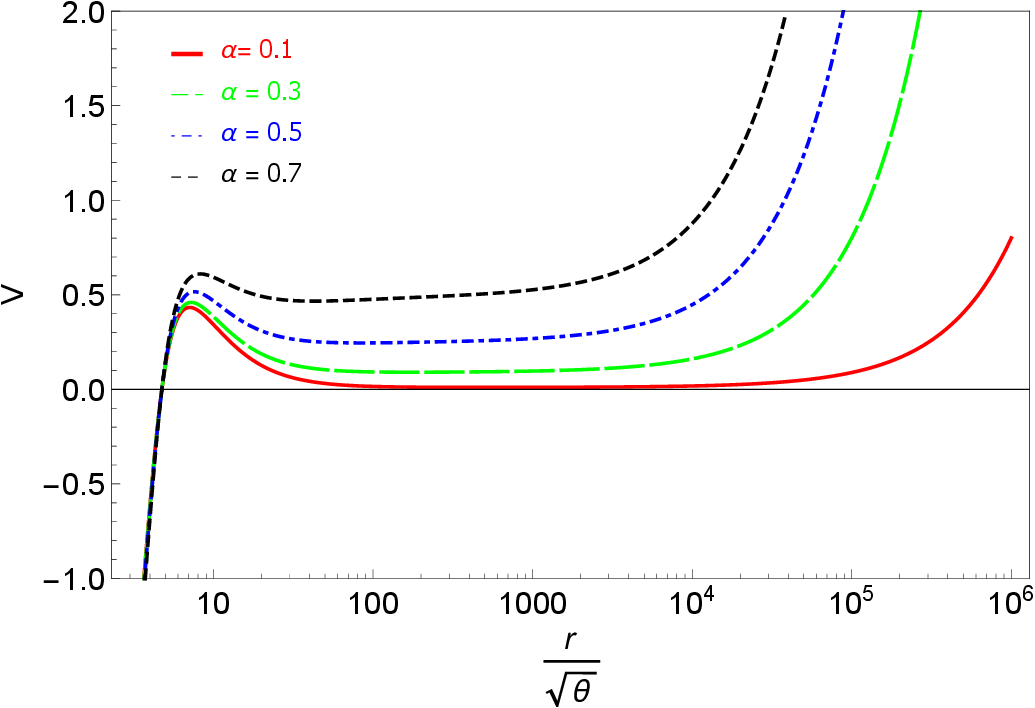}
        \caption{Left: effective potential in massless case $\alpha=0$ for various value of $\ell$ Right: effective potential in massive case $\alpha \neq 0$ for various value of $\alpha$ and fixed $\ell=2$. The other parameters are fixed to  $C=0.05,\,m^2c_1=0.01,\,m^2c_2=0.05,\theta=0.1$ and $M/\sqrt{\theta}=2.4.$ }
\label{fig:veff}
\end{figure}

Here we consider as massive scalar perturbation on spacetime background \eqref{line}. A scalar field propagating on curved spacetime is given by the Klein-Gordon equation
\begin{align}
    \nabla_{a}\nabla^a\psi - \alpha^2\psi &= 0,
\end{align}
where $\alpha^2$ is scalar field's mass squared. In spherical symmetric spacetime, scalar field $\psi$ can be separated into the following form
\begin{align}
    \psi &= e^{-i\omega t}\frac{R(r)}{r}Y(\Theta,\phi),
\end{align}
where $Y(\Theta,\phi)$ is spherical harmonic function. The scalar radial field equation can be recast into the Schr\"odinger-like form as
\begin{align}
   0 &= \frac{d^2R}{dr_\ast^2} + \left[\omega^2 - V \right]R , \\ 
    V &= e^{2\phi}\left( \alpha^2 + \frac{\ell(\ell+1)}{r^2} + \frac{2e^{2\phi}\phi'}{r} \right), \label{veff}
\end{align}
where prime denotes derivative with respect to $r$ and $\ell(\ell+1)$ is the eigenvalue of angular operator. The tortoise coordinate is defined such that
\begin{align}
    \frac{dr_{\ast}}{dr} &= e^{-2\phi},
\end{align}
It can be shown that in the massless case $\alpha=0$, the effective potential is finite i.e., $V\to C^2c_1m^4/4$ as $r\to \infty$. In contrast, $V$, in the massive case $\alpha \neq 0$, diverges as $r\to \infty$. In fact, the effective potential is monotonically increasing with $r$. These are demonstrated in Fig.~\ref{fig:veff}. Clearly, the maximum of effective potential increases with angular index $\ell$ as displayed in the left figure of Fig.~\ref{fig:veff}. With the chosen spacetime parameters i.e., $C=0.05,\,m^2c_1=0.01,\,m^2c_2=0.05$, $V \to 6.25\times 10^{-8}$ as $r\to \infty$. It is possible to render the $C^2c_1m^4/4$ non-zero finite value. Therefore, the $\alpha_{eff} \equiv C^2c_1m^4/4$ term effectively behaves as a scalar field mass. Thus, in the massless case $\alpha=0$, the appropriate boundary conditions leading to the quasinormal modes are $R\sim e^{\pm i\omega r_\ast}$ as $r_\ast \to \pm\infty$. In contrast, $V$ diverges asymptotically in the massive case as it is shown in the right figure of Fig.~\ref{fig:veff}. As scalar mass increases, the effective potential diverges more rapidly. The presence of mass term dramatically changes the potential's asymptotic structure. This requires special treatment when considering appropriate boundary conditions of the quasinormal modes. For the sake of simplicity, we shall particularly consider the massless case $\alpha=0.$ 

To compute the quasinormal frequencies, we implement Wentzel–Kramers–Brillouin (WKB) approximation method. This method is introduced to a study of black hole scattering problem \cite{Schutz:1985km}. Later, WKB method is extended to third \cite{Iyer:1986np} and sixth  order \cite{Konoplya:2003ii}. In \cite{Matyjasek:2017psv}, the WKB method is extended further to thirteenth order with the Pad\'e averaged. Several previous results can be reproduced with significant accuracy when implementing the Pad\'e averaged technique \cite{Matyjasek:2019eeu}. We note that, our computational calculation is based on the Wolfram's Mathematica code provided in \cite{Konoplya:2019hlu}.  In this work, we shall apply the third order WKB with the Pad\'e averaged to calculate the quasinormal frequencies of scalar field around noncommutative black hole in dRGT massive gravity. The WKB approximation works efficiently for $\ell>n$ when $n$ is overtone number as it is pointed out in \cite{Panotopoulos:2019qjk,Cardoso:2003vt}. In Table~\ref{Tab:QNMs}, we show example of quasinormal frequencies of massless scalar perturbation when varying $\ell$ and $n$. Remark that, the most fundamental mode ($\ell=n=0$) is also displayed but it might not be well-approximated by the WKB method. It can be seen that as $\ell$ increases, the real part of $\omega$ significantly increases while the imaginary part slightly decreases (in magnitude). On the contrary, the real part marginally decreases while the imaginary part notably increases (in magnitude) as the overtone number is larger. We note that for this point onward, only the most fundamental frequencies ($n=0$) are considered.

\begin{widetext}
\begin{table}[htb]
{\centering
\begin{center}
\caption{Quasinormal frequencies of massless scalar perturbation on noncommutative black hole in dRGT massive graivty with $C=0.05,\,m^2c_1=0.01,\,m^2c_2=0.05,M/\sqrt{\theta}=2.4$, and $\theta=0.1$. The frequencies are computed via the $3^{rd}$ order WKB with the Pad\'e averaged.}
\vspace{0.2cm}
\setlength{\tabcolsep}{7pt}
\begin{tabular}{ccccc}
\hline\\[-10pt]
$\ell$ &  $n=0$  & $n=1$ & $n=2$ & $n=3$  \\ \hline 

$0$ & $0.108428 - 0.045389i$ & $-$ & $-$ & $-$   \\

$1$ & $0.374373 - 0.040713 i$ & $-$ & $-$ & $-$ \\
      
$2$ & $0.630769 - 0.040309i$ & $0.627717 - 0.121194i$ & $-$ & $-$  \\

$3$ & $0.885705 - 0.040209i$ & $0.883507 - 0.120758i$ & $0.879193 - 0.201681 i$ & $-$   \\

$4$ & $1.140155 - 0.040169i$ & $1.138441 - 0.120585i$ & $1.135050 - 0.201229 i$ & $1.130058 - 0.282232i$  \\

\hline  
\end{tabular}
\label{Tab:QNMs}
\end{center}}
\end{table}
\end{widetext}

\begin{figure}[htbp]
    \centering
    \includegraphics[scale=.5]{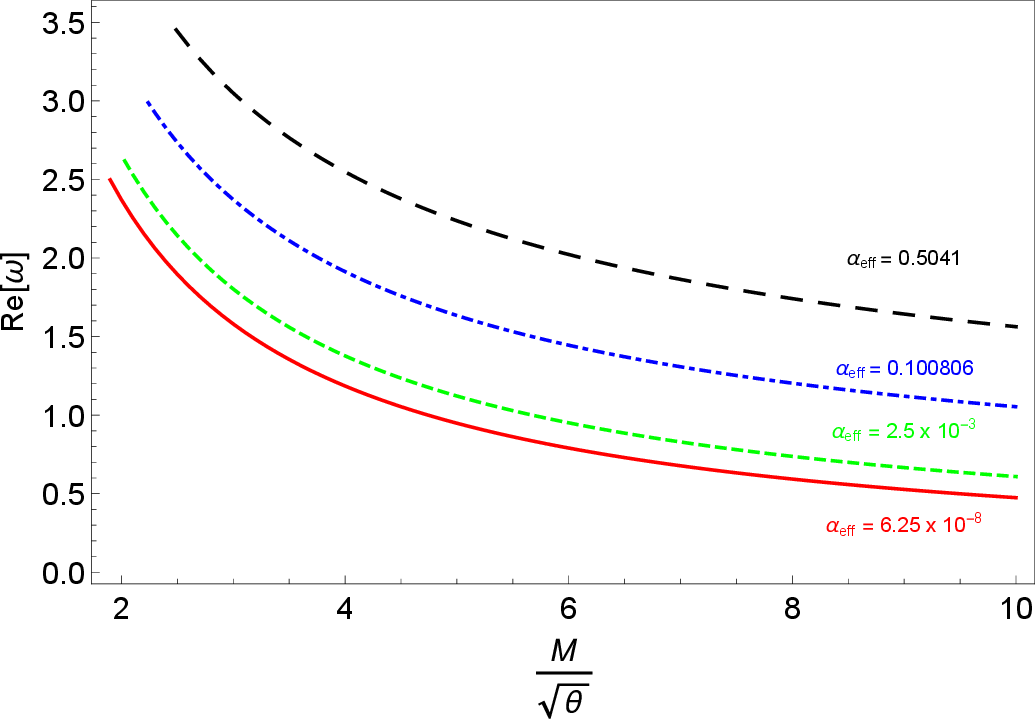}
    \includegraphics[scale=.5]{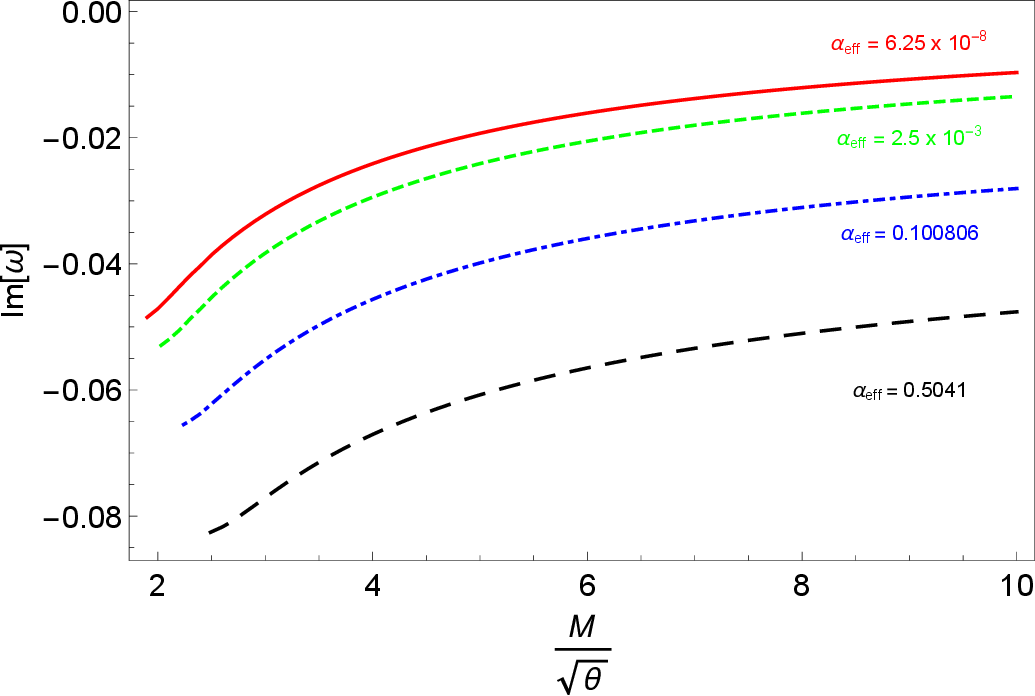}
        \caption{Quasinormal frequencies of noncommutative black hole in dRGT massive gravity against $M/\sqrt{\theta}$ for $\ell=2,\theta=0.01$ and $m^2c_2 =0.05$ for various value of $\alpha_{eff}.$}
        \label{fig:QNM1}
\end{figure}

Now we illustrate how $M/\sqrt{\theta}$ effect the quasinormal frequencies in Fig.~\ref{fig:QNM1}. In this plot, we pick four different values of effective scalar field's mass ($\alpha_{eff}$). This can be archived by adjusting $C$ and $c_1 = j/m^2$. For instance, $\alpha_{eff}=6.25\times 10^{-8}$ corresponds to $C=0.05,j=0.01$, $\alpha_{eff}=2.5\times 10^{-3}$ for  $C=1,j=0.1$, $\alpha_{eff}=0.100806$ for  $C=1.27,j=0.5$ and $\alpha_{eff}=0.5041$ for  $C=1.42,j=1$. It is worth-mentioning that, changing $C$ and $c_1$ leads to different asymptotic structure of the black hole metric \eqref{bhmetric}. From this figure, we find that as the mass increases, the real part of $\omega$ decreases. An opposite trends can be seen from the imaginary part (the right figure of Fig.~\ref{fig:QNM1}). The $Im(\omega)$ becomes less negative as $M/\sqrt{\theta}$ increases. We also observe that the higher the $\alpha_{eff}$, the lower $Re(\omega)$ and $Im(\omega)$. Remark that in each curve, the smallest $M/\sqrt{\theta}$ indicates the black hole that is very close to its extremal limit.

\begin{figure}[htbp]
    \centering
    \includegraphics[scale=.5]{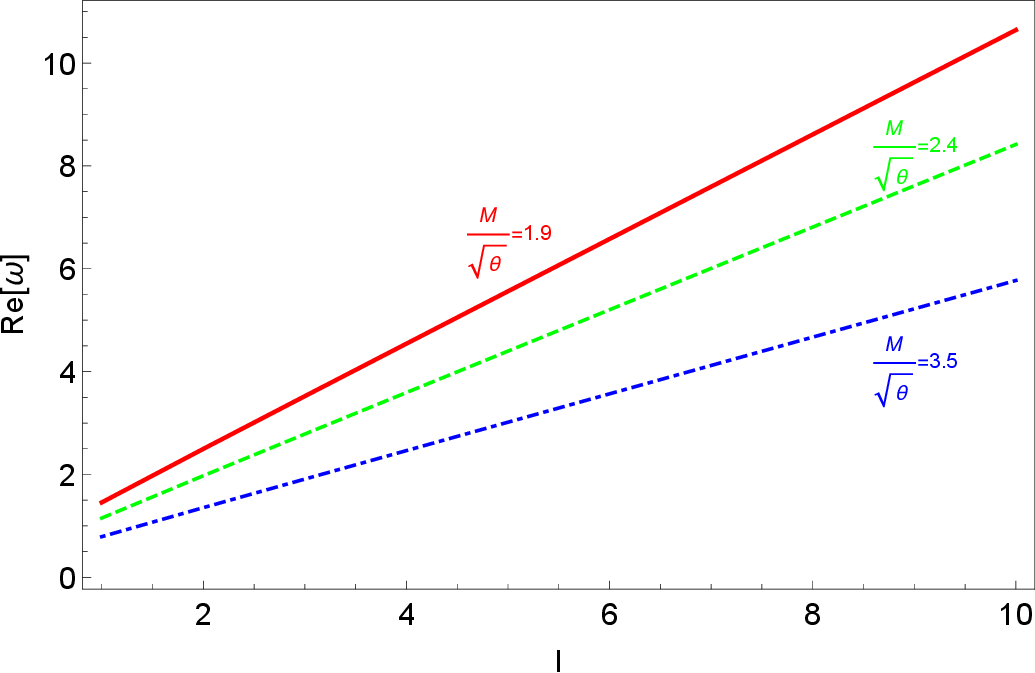}
    \includegraphics[scale=.5]{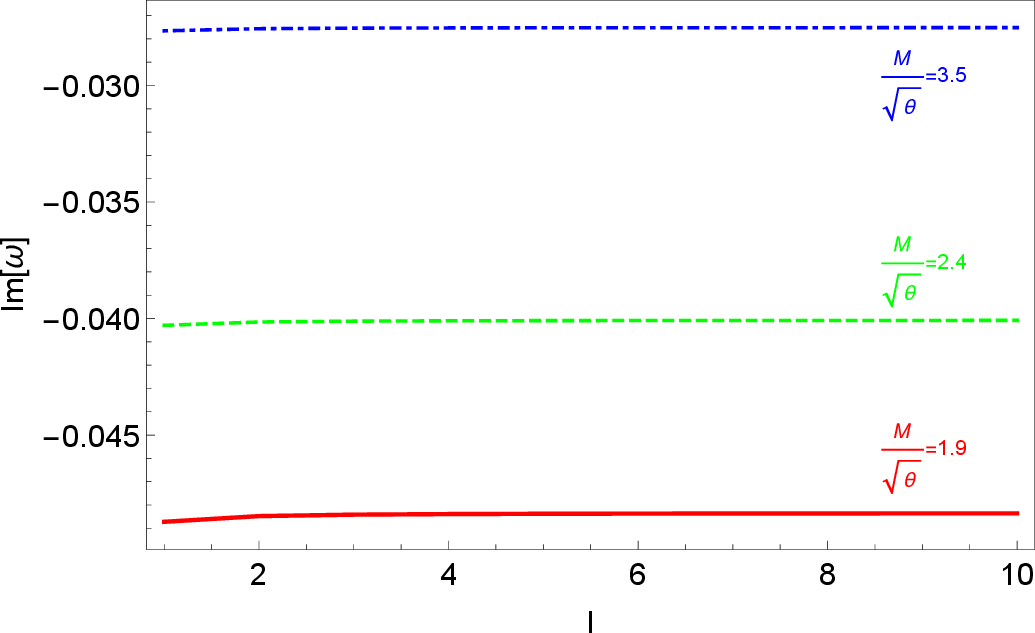}
    \caption{Quasinormal frequencies of noncommutative black hole in dRGT massive gravity against $\ell$ for $\theta=0.01,m^2c_2 =0.05$ and $\alpha_{eff}=6.25\times 10^{-8}$ for three distinct values of $M/\sqrt{\theta}.$}
    \label{fig:QNM2}
\end{figure}

In Fig.~\ref{fig:QNM2}, quasinormal frequencies as function of angular index $\ell$ are displayed with three different values of $M/\sqrt{\theta}$. The angular index $\ell$ affects the $Re(\omega)$ much more significantly than the $Im(\omega)$. The real part increases monotonically with $\ell$. In contrast, the imaginary part narrowly increases with $\ell$. For $M/\sqrt{\theta}=1.9$, the increasing of the $Im(\omega)$ is noticeable at small $\ell$. However as $M/\sqrt{\theta}$ increases, this behaviour becomes less obvious. Lastly, with smaller $M/\sqrt{\theta}$, the rate at which the $Re(\omega)$ changes with $\ell$ is higher than those of large $M/\sqrt{\theta}$.

\begin{figure}[htbp]
    \centering
    \includegraphics[scale=.5]{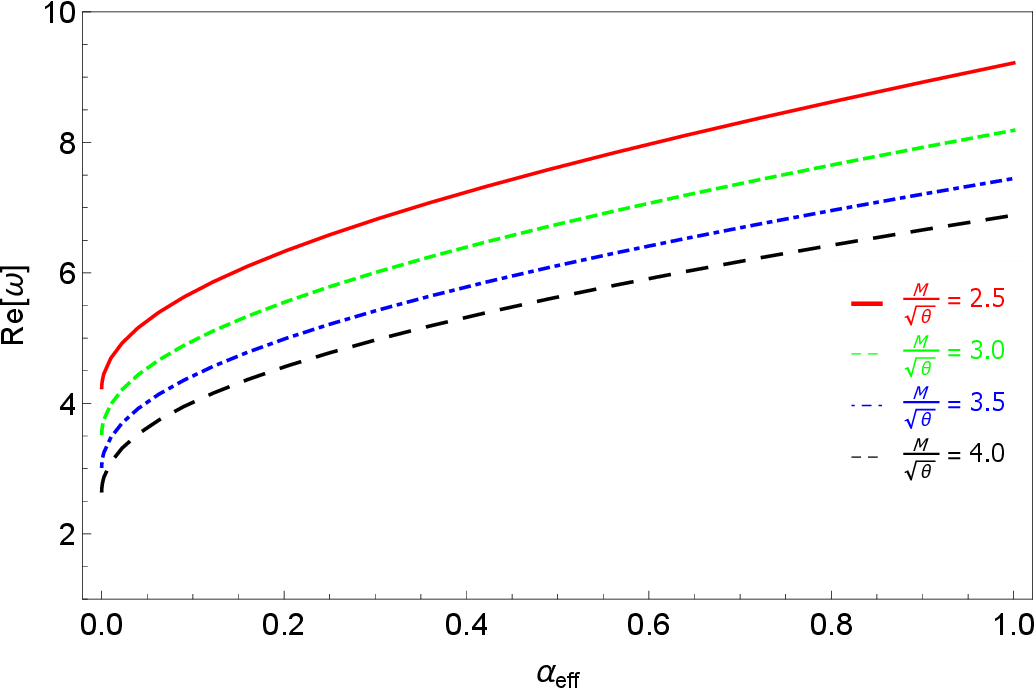}
    \includegraphics[scale=.5]{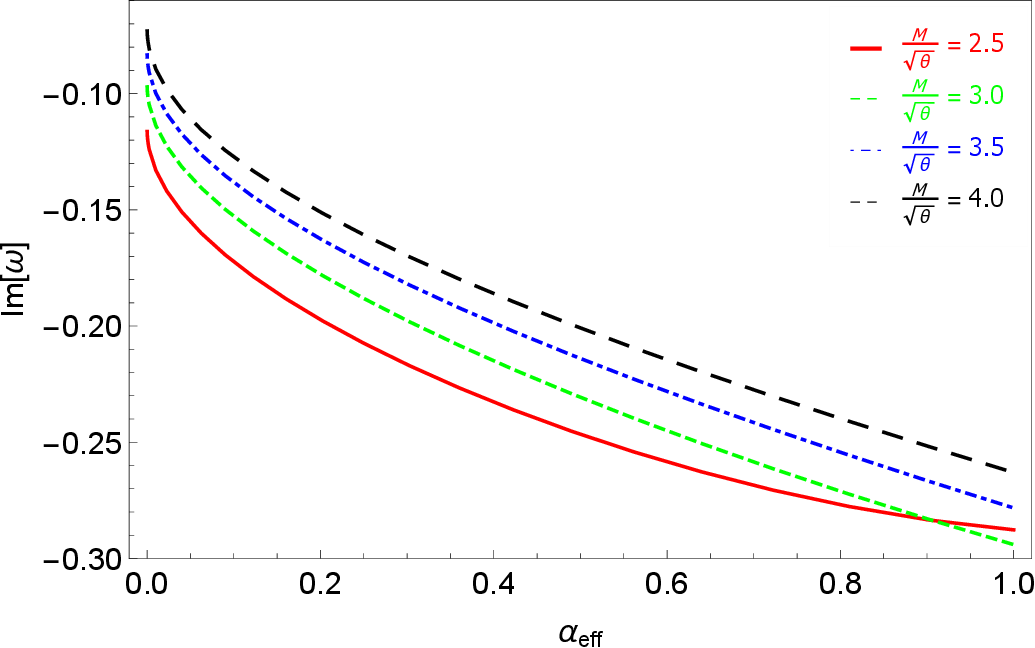}
\caption{Quasinormal frequencies of noncommutative black hole in dRGT massive gravity against $\alpha_{eff}$ for $\theta=0.01,m^2c_2 =0.05$ and $\ell=5$ for four distinct values of $M/\sqrt{\theta}.$}
\label{fig:QNM3}
\end{figure}

In addition, we show quasinormal frequencies as function of effective scalar's mass $\alpha_{eff}$ in Fig.~\ref{fig:QNM3}.
The real part of $\omega$ is larger as $\alpha_{eff}$ increases. Remarkably, the imaginary part becomes more negative as $\alpha_{eff}$ is larger. This is in opposition to many previous results that increasing scalar field mass tends to slower the decay \cite{Simone:1991wn,Konoplya:2002wt,Konoplya:2004wg}. It is important to emphasize that $\alpha_{eff}$ is not an actual scalar field mass but rather play a similar role as $\alpha^2$ in the effective potential \eqref{veff}. Moreover, the presence of $\alpha_{eff}$ directly relates to the spacetime parameters $C$ and $c_1$. Therefore, non-zero value of $\alpha_{eff}$ inevitably affects the spacetime's geometry.

\subsection{The eikonal limit}

The WKB approximation technique is considerably more accurate in the eikonal limit $\ell \to \infty$. Under this limit, the effective potential \eqref{veff} is reduced to the following form
\begin{align}
    V_{eik} &= e^{2\phi}\frac{\ell^2 }{r^2}.
\end{align}
With the simplified effective potential, it is possible to express quasinormal frequency in an analytic form \cite{Cardoso:2008bp}
\begin{align}
    \omega_{eik} &= \Omega \ell - i \left(n + \frac{1}{2}\right)|\lambda_L|, \label{weik}
\end{align}
where the coordinate angular velocity $\Omega$ at the unstable null orbit and the Lyapunov exponent $\lambda$ are given by
\begin{align}
    \Omega &= \left. \sqrt{ \frac{e^{2\phi}}{r^2}} \right \vert_{r=r_0}, \\
    \lambda_L &= \frac{1}{\sqrt{2}}\left.\sqrt{ \frac{r^2}{e^{2\phi}} \frac{d^2}{dr_\ast^2}\left(\frac{e^{2\phi}}{r^2}\right) } \right \vert_{r=r_0},
\end{align}
where $\Omega$ and $|\lambda_L|$ are evaluated at $r=r_0$ i.e., the peak of the potential. More interestingly, there is a connection between the real part of eikonal quaisnormal modes and black hole shadow radius \cite{Jusufi:2019ltj,Jusufi:2020dhz,Jusufi:2020mmy}. It is suggests that \cite{Jusufi:2019ltj}
\begin{align}
    Re(\omega_{eik}) &= \lim_{\ell \gg 1} \frac{\ell}{R_s}, \label{eik&shadow}
\end{align}
where $R_s$ is the shadow radius. We will come back to this point when discussing optical property in the following section.

\begin{figure}[htbp]
    \centering
    \includegraphics[scale=.5]{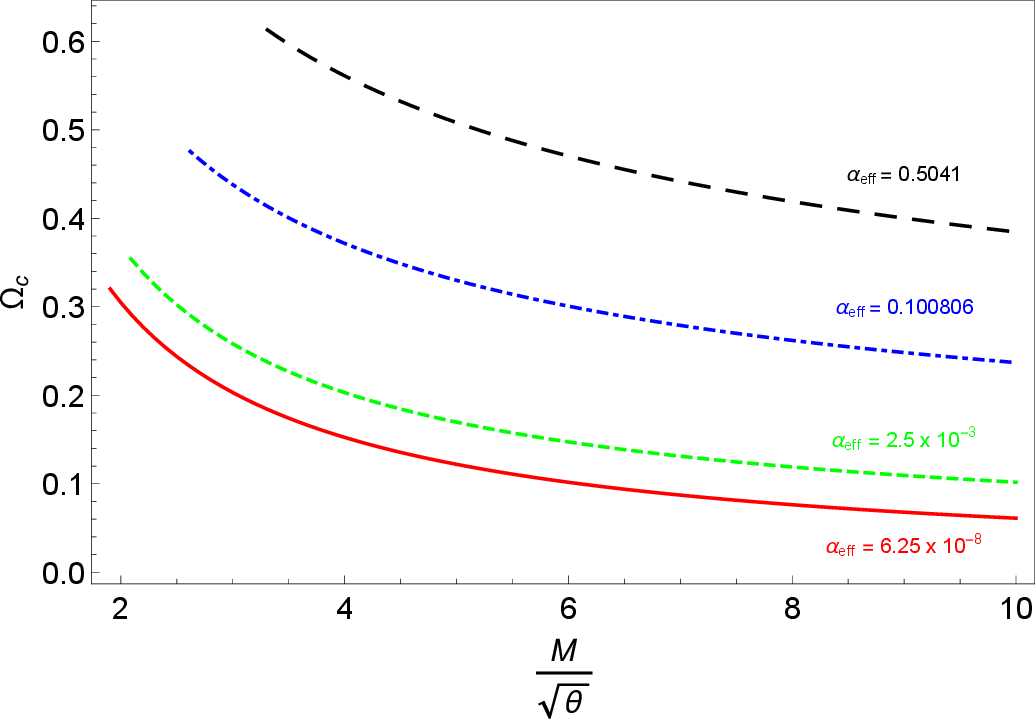}
    \includegraphics[scale=.5]{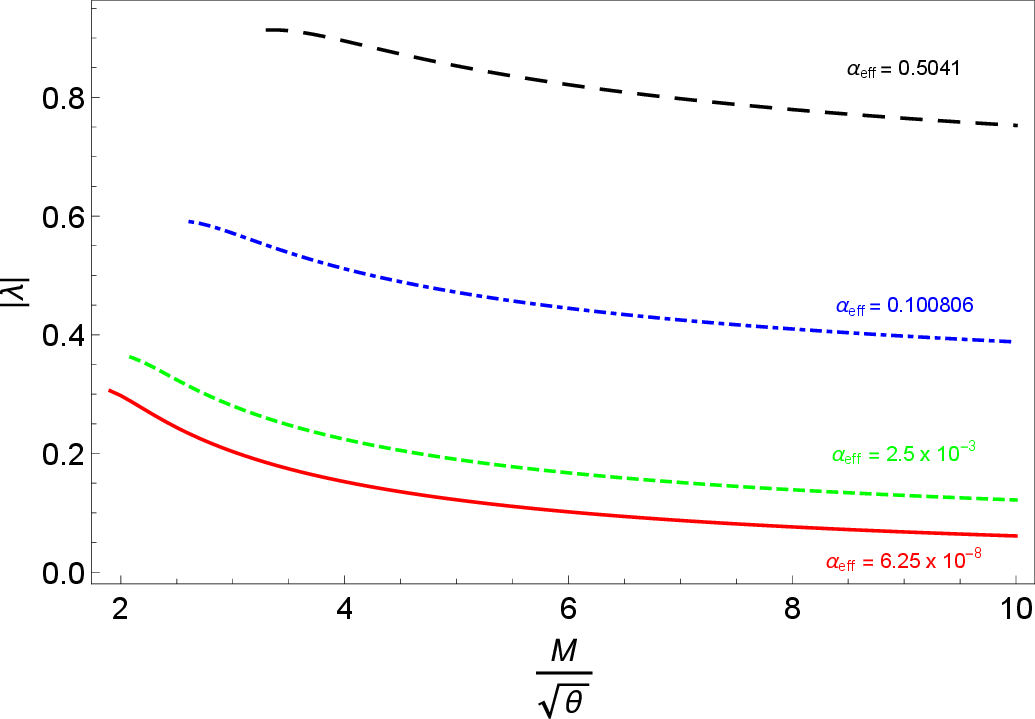}
        \caption{Coordinate angular velocity $\Omega$ (left) and the Lyapunov exponent $|\lambda|$ (right) plotted against $M/\sqrt{\theta}$ for $\theta=0.1$ with four chosen values of $\alpha_{eff}.$}
    \label{fig:eik}
\end{figure}

Now let's explore the behaviour of coordinate angular velocity and the Lyapunov exponent under influence of $M/\sqrt{\theta}$ in Fig.~\ref{fig:eik}. Clearly, both $\Omega$ and $|\lambda|$ is smaller as the black hole mass increases. Moreover, the higher the effective scalar field's mass, the larger angular velocity and the Lyapunov exponent.

\section{Optical Property: Shadow}\label{sec:shadow}

This segment investigates the enigmatic shadows produced by black holes—shadowy regions in space-time covering the event horizon of these astronomical occurrences. Black holes are distinguished by an incredibly potent gravitational force, so formidable that light cannot escape once it crosses the event horizon \cite{EslamPanah:2020hoj, Konoplya2019}. Consequently, a mysterious and darkened circle emerges around the periphery of the black hole, standing out against the backdrop of adjacent matter or light. The size and configuration of this obscure silhouette offer valuable insights into the characteristics of black holes and the essence of gravity. Recent progress in astronomy and technology now enables the acquisition of images depicting these elusive black hole shadows, markedly advancing our comprehension of these puzzling celestial entities \cite{gogoi_joulethomson_2023, Gogoi:2023ffh, Gogoi:2024vcx}.

To obtain the shadow of the black hole, we begin with the standard Euler-Lagrange equation as given below:
\begin{equation}\label{shadow1}
\frac{d}{d\tau}\!\left(\frac{\partial\mathcal{L}}{\partial\dot{x}^{\mu}}\right)-\frac{\partial\mathcal{L}}{\partial x^{\mu}}=0,
\end{equation}
where the term $\mathcal{L}$ represents the Lagrangian of the system and it can be given by,
\begin{equation}\label{shadow11}
\mathcal{L}(x,\dot{x})=\frac{1}{2}\,g_{\mu\nu}\dot{x}^{\mu}\dot{x}^{\nu}.
\end{equation}
In this investigation, we have considered a static and spherically symmetric spacetime metric in the presence of noncommutative geometry and hence the Lagrangian can be further written as
\begin{equation}\label{shadow2}
\mathcal{L}(x,\dot{x})=\frac{1}{2}\left[-h(r)\,\dot{t}^{2}+\frac{1}{h(r)}\,\dot{r}^{2}+r^{2}\left(\dot{\Theta}^{2}+\sin^{2}\Theta\dot{\phi}^{2}\right)\right].
\end{equation}
One may note that the dot over the variables in the above expression stands for derivative with respect to the proper time $\tau$.

When we select the equatorial plane, where the polar angle $\Theta$ is fixed at $\pi/2$, we can determine the conserved energy $\mathcal{E}$ and angular momentum $L$ of a particle using the killing vectors $\partial/\partial \tau$ and $\partial/\partial \phi$. This yields the expressions
\begin{equation}
\mathcal{E}=h(r)\,\dot{t},\quad L=r^{2}\dot{\phi},
\end{equation}
where $h(r)$ is nothing but the metric function of the black hole previously denoted by $e^{2 \phi}$. It is a function of the radial coordinate $r$. In the above expression, $\dot{t}$ and $\dot{\phi}$ are the time and azimuthal components of the particle's four-velocity.

For photons, which travel along null geodesics in curved spacetime, the geodesic equation on the equatorial plane takes the form
\begin{equation}\label{eq22}
-h(r)\,\dot{t}^{2}+\frac{\dot{r}^{2}}{h(r)}\,+r^{2}\dot{\phi}^{2} = 0.
\end{equation}
By combining Eq.\ \eqref{eq22} with the conserved quantities $\mathcal{E}$ and $L$, we can derive the orbital equation for photons which is given by \cite{gogoi_joulethomson_2023, Gogoi:2024vcx}:
\begin{equation}\label{eff}
\left(\frac{dr}{d\phi}\right)^{2}=V_{eff},
\end{equation}
where the effective potential $V_{eff}$ has the following explicit form:
\begin{equation}
V_{eff}= r^{4} \left[\frac{\mathcal{E}^{2}}{L^{2}}-\frac{h(r)}{r^{2}}\right].
\end{equation}
This effective potential helps us to obtain the photon sphere position around the black hole. To obtain the same, we first try to reduce the potential expression defined above.
In a more convenient radial form, using a reduced form of potential, Eq.\ \eqref{eff} becomes
\begin{equation} \label{redpot}
    V_r(r) = \dfrac{1}{\xi^2} - \dot{r}^2/L^2,
\end{equation}
where $\xi$ is the impact parameter defined as $\xi=L/\mathcal{E}$. The reduced form of the potential is denoted by $V_r(r)$ and it has the following explicit form:
\begin{equation}\label{pot}
V_r(r) = \frac{h(r)}{ r^2}.
\end{equation}

To study the shadows of black holes, we need to focus on a specific point along the trajectory, denoted as $r_{ph}$, which corresponds to the turning point. This point marks the location of the light ring that encircles the black hole and is also the radius of the photon sphere, a key element in analyzing the black hole's shadow. At this turning point, the effective potential $V_{eff}$ as well as $\dot{r}$ vanishes, indicating the critical impact parameter $\xi_{crit}$ \cite{pantig_shadow_2022, Ovgun:2018tua, Papnoi:2014aaa, Ovgun:2019jdo}:
\begin{equation}\label{cond01}
\left.V_{eff}\right|_{r_{ph}} = 0,
\end{equation}
which can be expressed as
\begin{equation}\label{impact}
\frac{1}{\xi_{crit}^{2}}=\frac{h(r_{ph})}{r_{ph}^{2}}.
\end{equation}
One may note that this same expression also can be obtained by using $\dot{r} = 0$ at $r = r_{ph}$ in the equation of reduced potential Eq. \eqref{redpot}. 

The radius of the photon sphere can be obtained from 
\begin{equation}
\left.V'_{eff}\right|_{r_{ph}} = 0,
\end{equation}
where the prime represents derivative w.r.t. $r$. This condition can be further simplified by using Eq. \eqref{cond01}. Similarly, one can also use the expression of reduced potential to obtain the photon sphere radius by calculating the extremum position. It shows that the radius of the photon sphere $r_{ph}$ can be determined by finding the extremum of the function $\mathcal{A}(r)$:
\begin{equation}
\left.\frac{d}{dr}\,\mathcal{A}(r)\right|_{r_{ph}} = 0,
\end{equation}
which, in turn, can be rewritten as
\begin{equation}
\frac{h^{\prime}(r_{ph})}{h(r_{ph})}-\frac{\mathcal{Q}^{\prime}(r_{ph})}{\mathcal{Q}(r_{ph})}=0,
\end{equation}
where $\mathcal{A}(r)=\mathcal{Q}(r)/h(r)$ and $\mathcal{Q}(r)=r^{2}$. This critical radius $r_{ph}$ and its corresponding impact parameter provide crucial insights into the nature of black hole shadows.

To determine the shadow of a black hole, we can rewrite Eq.\ \eqref{eff} with Eq.\ \eqref{impact} in terms of the function $\mathcal{A}(r)$ as
\begin{equation}
\left(\frac{dr}{d\phi}\right)^{\!2}= \mathcal{Q}(r)h(r)\left(\frac{\mathcal{A}(r)}{\mathcal{A}(r_{ph})}-1\right).
\label{eq33}
\end{equation}
This equation allows us to calculate the shadow radius of the black hole. When considering a scenario where a static observer is positioned at a distance $r_0$ from the black hole, we can find the angle $\alpha$ between the light rays emitted by the observer and the radial direction of the photon sphere. This angle is given by
\begin{equation}
\cot\alpha=\frac{1}{\sqrt{h(r)\mathcal{Q}(r)}}\left.\frac{dr}{d\phi}\right|_{r\,=\,r_{0}}\!\!\!\!\!\!\!\!\!\!\!.
\end{equation}
Combining this with Eq.\ \eqref{eq33}, we get
\begin{equation}
\cot^{2}\!\alpha=\frac{\mathcal{A}(r_{0})}{\mathcal{A}(r_{ph})}-1,
\end{equation}
which can be rewritten using the relation $\sin^{2}\!\alpha=1/(1+\cot^{2}\!\alpha)$ as
\begin{equation}
\sin^{2}\!\alpha=\frac{\mathcal{A}(r_{ph})}{\mathcal{A}(r_{0})}.
\end{equation}
The above equation helps us determine the angular size of the black hole's shadow as seen by an observer at a distance $r_0$ from the black hole.

By substituting the expression for $\mathcal{A}(r_{ph})$ from Eq.\ \eqref{impact} and $\mathcal{A}(r_{0}) = r_0^2/h(r_0)$, the black hole's shadow radius for a static observer at $r_{0}$ is estimated as \cite{gogoi_joulethomson_2023}:
\begin{equation}\label{shadoweq}
R_{s}=r_{0}\sin\alpha=\sqrt{\frac{r_{ph}^2h(r_{0})}{h\left(r_{ph}\right)}}.
\end{equation}
In the limit of asymptotically flat spacetime as $r_0 \rightarrow \infty$, which corresponds to a static observer at a large distance where $h(r_0) \rightarrow 1$, the shadow radius $R_s$ simplifies to:
\begin{equation}
R_{s} = \frac{r_{ph}}{\sqrt{h(r_{ph})}}.
\end{equation}
This expression gives the shadow radius of the black hole as observed by a distant static observer for asymptotically flat spacetime. Clearly, the right hand side of the above equation is the inverse of angular velocity i.e., $1/\Omega$. Hence, it is in agreement with \eqref{eik&shadow} discussed earlier. However, it is important to mention that the black hole solution being investigated in this work is not asymptotically flat and hence with a variation in $r_0$, even at a very far distance away, can have impacts on the black hole shadow. Due to this nature, we have thoroughly investigated the impacts of $r_0$ on the black hole shadow for different sets of the model parameters. 

\begin{figure*}[h!]
      	\centering{
      	\includegraphics[scale=0.6]{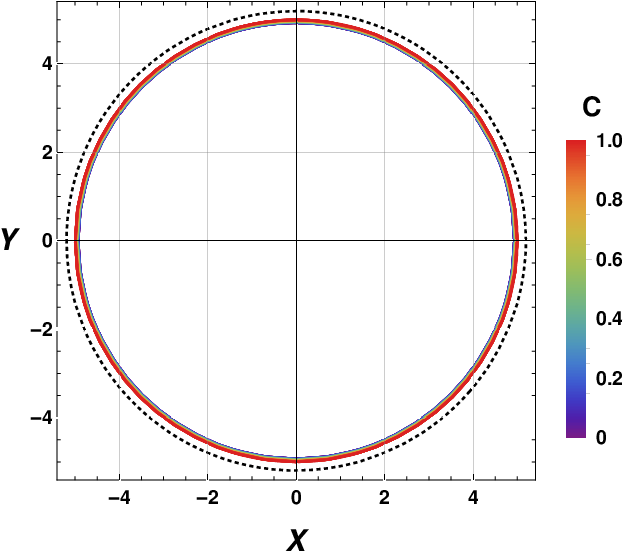} \hspace{1cm}
       \includegraphics[scale=0.6]{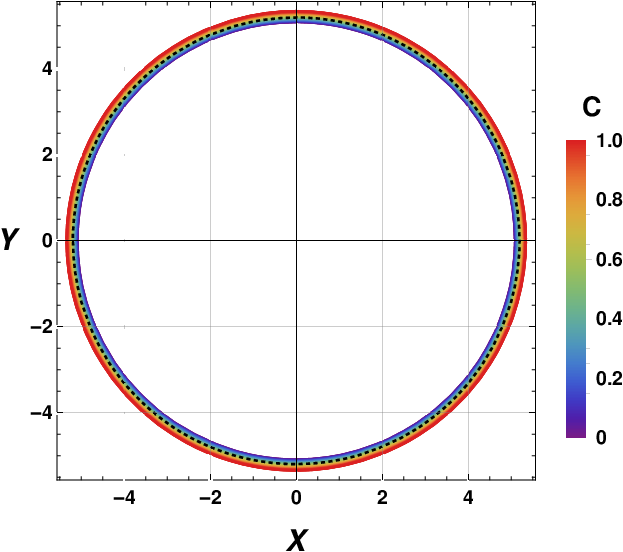}}
      	\caption{Stereographic projection of black hole shadow using $M = 1, c_1 = 0.5, c_2 = 0.15, m = 0.1$ and $ \theta = 0.2$. On the first panel, we have used $ r_0 = 20$ and on the second panel $r_0 = 50$. The black dotted circle represents Schwarzschild's black hole shadow. }
      	\label{figSh01}
      \end{figure*}

      \begin{figure*}[h!]
      	\centering{
      	\includegraphics[scale=0.6]{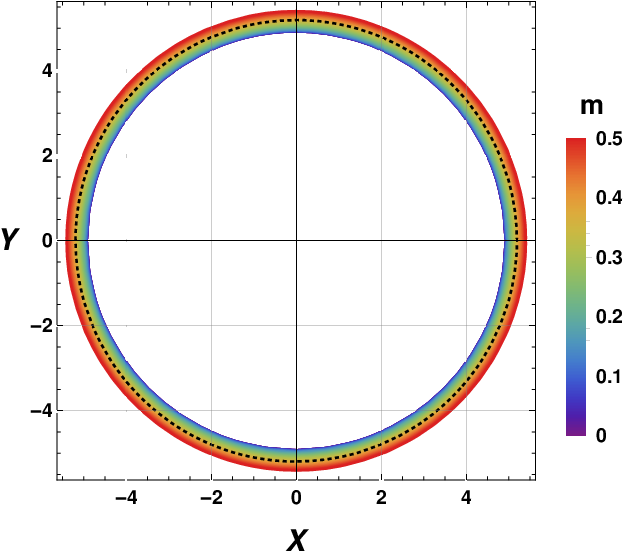} \hspace{1cm}
       \includegraphics[scale=0.6]{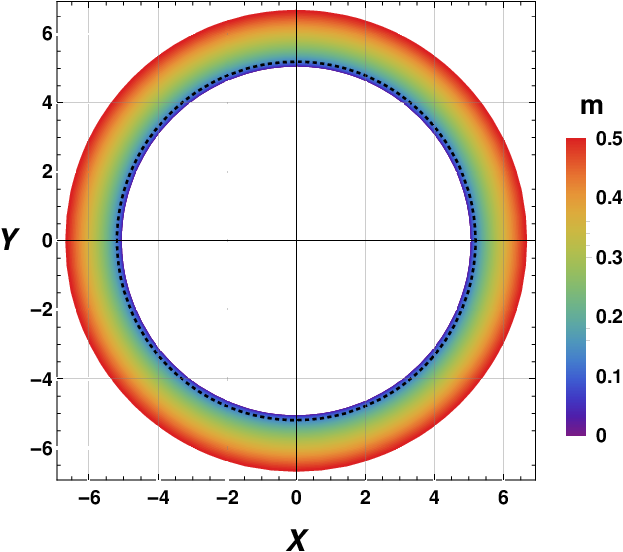}}
      	\caption{Stereographic projection of black hole shadow using $M = 1, C = 0.5, c_1 = 0.35, c_2 = 0.2$ and $ \theta = 0.2$. On the first panel, we have used $r_0 = 20$ and on the second panel $r_0 = 50$. The black dotted circle represents Schwarzschild's black hole shadow. }
      	\label{figSh02}
      \end{figure*}
      \begin{figure*}[h!]
      	\centering{
      	\includegraphics[scale=0.6]{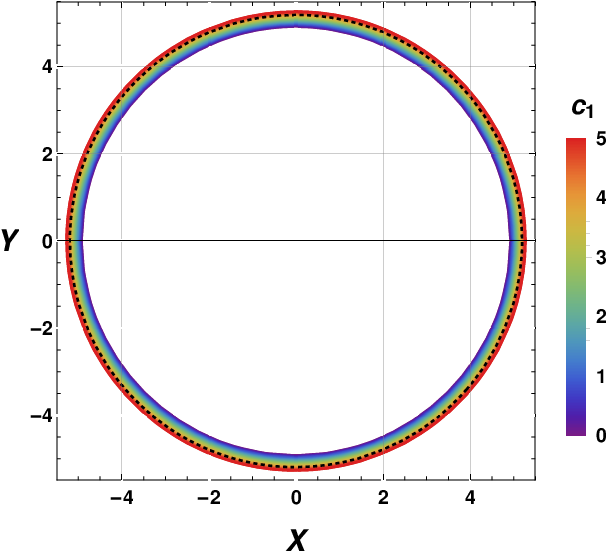} \hspace{1cm}
       \includegraphics[scale=0.6]{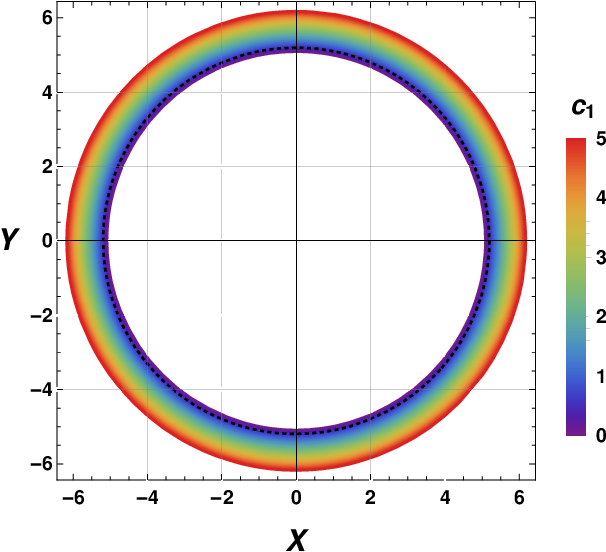}}
      	\caption{Stereographic projection of black hole shadow using $M = 1, C = 0.5, c_2 = 0.15, m = 0.1 $ and $ \theta =0.2$. On the first panel, we have used $r_0 = 20$ and on the second panel $r_0 = 50$. The black dotted circle represents Schwarzschild's black hole shadow. }
      	\label{figSh03}
      \end{figure*}
      \begin{figure*}[h!]
      	\centering{
      	\includegraphics[scale=0.6]{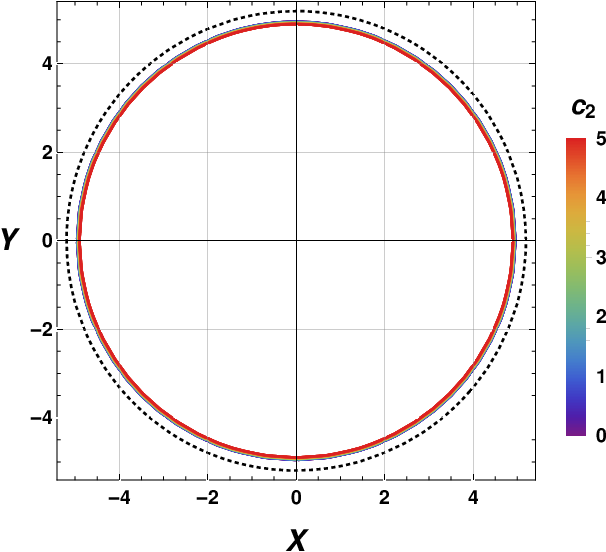} \hspace{1cm}
       \includegraphics[scale=0.6]{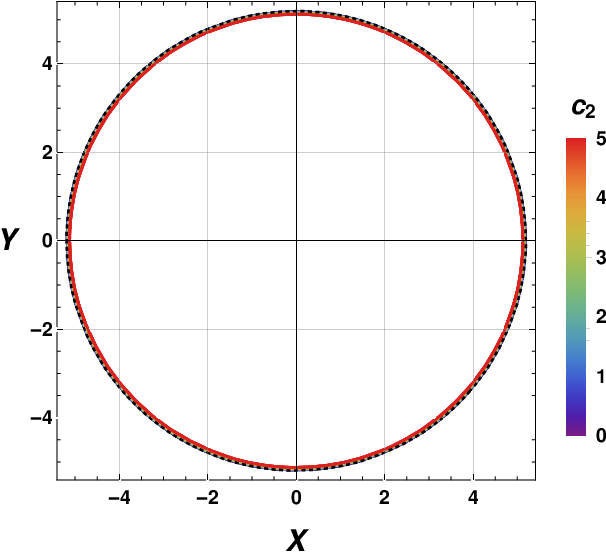}}
      	\caption{Stereographic projection of black hole shadow using $M = 1, C = 0.5, c_1 = 0.35, m = 0.1 $ and $ \theta = 0.2$. On the first panel, we have used $r_0 = 20$ and on the second panel $r_0 = 50$. The black dotted circle represents Schwarzschild's black hole shadow. }
      	\label{figSh04}
      \end{figure*}

      \begin{figure*}[h!]
      	\centering{
      	\includegraphics[scale=0.6]{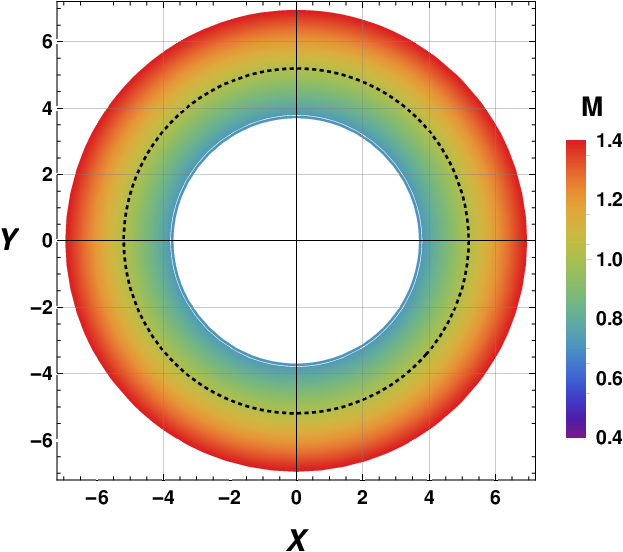} \hspace{1cm}
       \includegraphics[scale=0.6]{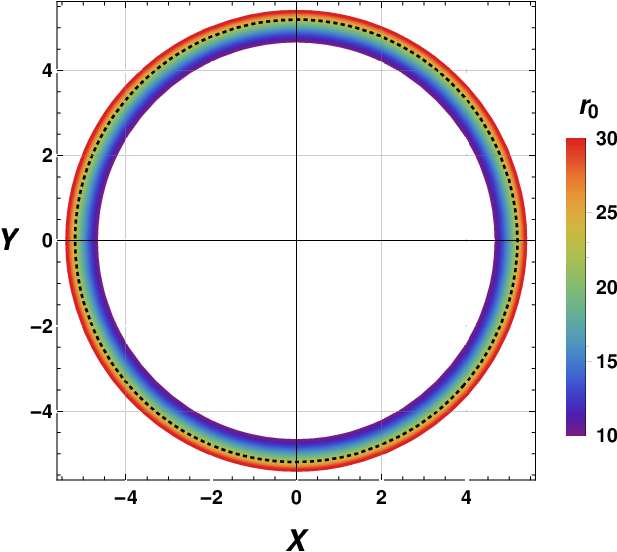}}
      	\caption{Stereographic projection of black hole shadow using $C = 0.5, c_1 = 0.35, c_2 = 0.2, m = 0.3 $ and $ \theta = 0.1$. On the left panel, we have used $r_0 = 20$ and on the right panel $M = 1$. The black dotted circle represents Schwarzschild's black hole shadow.  }
      	\label{figSh05}
      \end{figure*}

      \begin{figure*}[h!]
      	\centering{
      	\includegraphics[scale=0.6]{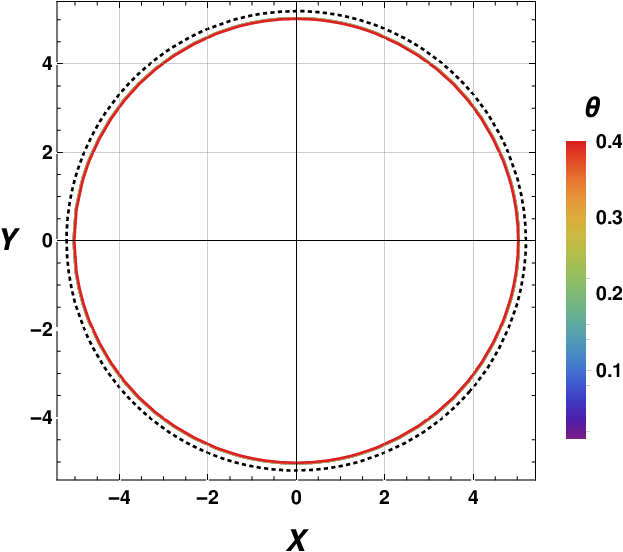} \hspace{1cm}
       \includegraphics[scale=0.6]{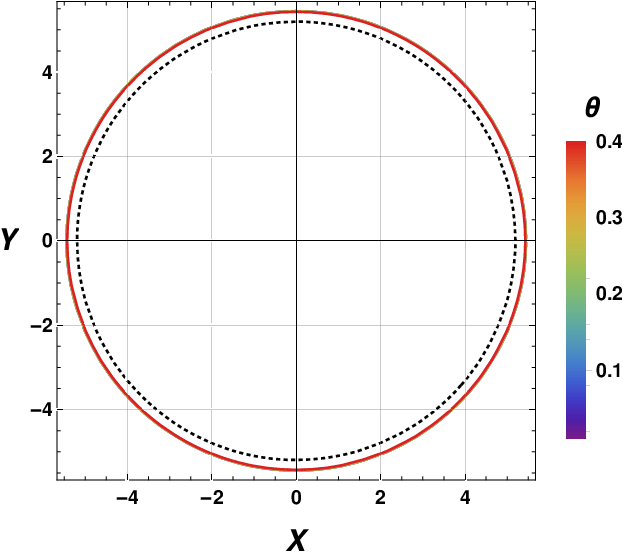}}
      	\caption{Stereographic projection of black hole shadow using $C = 0.5, c_1 = 0.35, c_2 = 0.1, m = 0.2$ and $ M = 1 $. On the left panel, we have used $r_0 = 20$ and on the right panel $r_0 = 50$. The black dotted circle represents Schwarzschild's black hole shadow.  }
      	\label{figSh06}
      \end{figure*}
\begin{figure*}[h!]
      	\centering{
      	\includegraphics[scale=0.75]{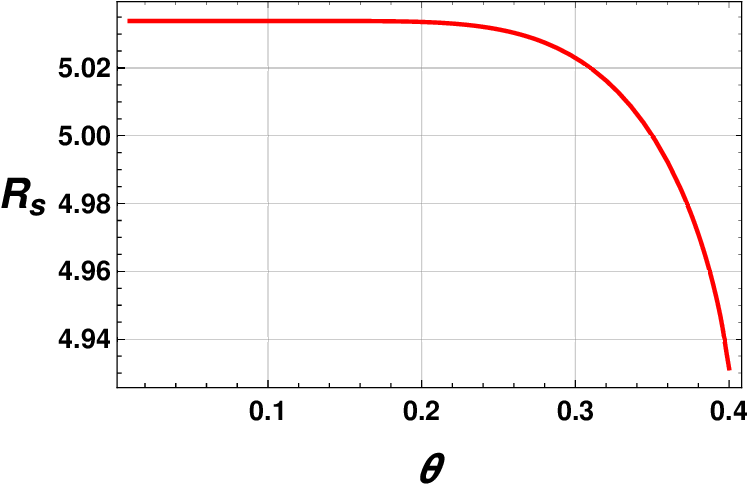}}
      	\caption{Variation of black hole shadow w.r.t $\theta$ at observer distance $r_0 = 20$ using $C = 0.5, c_1 = 0.35, c_2 = 0.1, m = 0.2$ and $ M = 1 $. }
      	\label{figSh07}
      \end{figure*}

In Fig. \ref{figSh01}, we have shown the variation of black hole shadow for different values of model parameter $C$. On the left panel, we choose observer distance $r_0 = 20$ and on the right panel, $r_0 = 50$. One may observe that with an increase in the value of the model parameter $C$, the size of the black hole shadow increases slowly. This variation is, however, very small. The black dotted circle in these figures represent the shadow radius of a standard Schwarzschild black hole. It is also seen that an increase in $r_0$ increases the shadow size of the black hole and at the same time, the impact of the model parameter also becomes more significant at larger $r_0$ values. So, the impacts of such a model parameter can be effectively constrained using observational data.
In Fig. \ref{figSh02}, we have shown the impacts on the black hole shadow by the parameter $m$. Similar to the scenario of $C$, we observe an increase in the shadow radius with an increase in the model parameter $m$ value. However, the impacts of $m$ on the black hole shadow are more significant. Similar to the previous case, at far distance away from the black hole, the impacts of $m$ are more clearly visible. 
In Fig. \ref{figSh03}, we have shown the effects of parameter $c_1$ on the black hole shadow. On the left panel, we have used the observer at $r_0 = 20$ and on the right panel, we have used the observer at $r_0 = 50$. One may note that, at a far distance away from the black hole, the impact of the model parameter $c_1$ is more significant. In both cases, with an increase in the model parameter $c_1$, we observe an increase in the shadow size of the black hole.
From Fig. \ref{figSh04}, one may observe that $c_2$ has a very small impact on the black hole shadow. The impact is also almost negligible with a variation of the observer's distance. Unlike the previous scenarios, here, an increase in $c_2$ decreases the black hole shadow size slowly.
In Fig. \ref{figSh05}, impacts of black hole mass $M$ and observer distance $r_0$ on the black hole shadow are shown. Mass $M$ has a usual impact on the shadow of the black hole. With an increase in the observer distance $r_0$, the shadow size of the black hole increases significantly.
Impacts of the noncommutative parameter $\theta$ on the black hole shadow are shown in Figs.~\ref{figSh06} and \ref{figSh07}. From Fig. \ref{figSh06}, one may note that the impact of $\theta$ on the black hole shadow is very small and it is also almost negligible with a variation of the observer's distance from the black hole. Since noncommutative parameter is an important part of our investigation, we further plot Fig. \ref{figSh07} to get a clearer idea of the impact of $\theta$ on the shadow radius of the black hole. As predicted earlier, the impact of $\theta$ on the black hole shadow radius is found to be very small. As expected $\theta$ impacts the black hole shadow nonlinearly and with an increase in $\theta$, the black hole shadow radius decreases as shown in Fig. \ref{figSh07}.

\section{Concluding Remarks}\label{sec06}
In the context of de Rham-Gabadadze-Tolley-like massive gravity, we have examined the properties and dynamics of the noncommutative black hole with Gaussian distribution in this article. Through our investigation, we were able to determine the relationship between the graviton mass m and the metric potential, temperature, entropy, heat capacity, and free energy. To explain various physical features of our model, we take into consideration a particular set of values for the model parameters. Fig.~\ref{fig1} shows that, for a particular set of model parameters, the non-commutative black hole has a minimum mass ($M_{r_H}$) below which it does not exist. For a given BH mass $M=M_{r_H}$ (where $M_{r_H}$ is the critical mass), there is one event horizon, and for $M>M_{r_H}$, there are two event horizons. Notably, we found that despite the various values of $\frac{M}{\sqrt{\theta}}$, we observed for the particular Gaussian form of density, the minimal horizon is always located at the distance $r=2.9\sqrt{\theta}$. Our research indicates that the temperature of the non-commutative black hole disappears as the horizon radius approaches the minimal horizon. The weak energy condition is always satisfied for our model but for $r>2.9 \sqrt{\theta}$, it is noteworthy that the strong energy condition breaks down.

A linear scalar perturbation on the noncommutative black hole in dRGT massive gravity is explored. We find that the model parameters $C,m$ and $c_1$ combined effectively as scalar field's mass ($\alpha_{eff}$). We have seen that the real part of frequency decreases with $M/\sqrt{\theta}$. In contrast, the real part of frequency increases with $\ell$ and $\alpha_{eff}$. In addition, the imaginary part of frequency becomes less negative as $M/\sqrt{\theta}$ is larger. We have also observed that increasing $\ell$ leads to slightly increasing $Im(\omega)$. Remarkably, the $Im(\omega)$ is more negative with increasing in $\alpha_{eff}$. Additionally, we investigate the QNMs in the eikonal limit where the angular velocity $\Omega$ and the Lyapunov exponent $\lambda$ are shown to be decreasing function with $M/\sqrt{\theta}$.

We have also investigated the shadow of the noncommutative black hole in this investigation. We have seen that the shadow radius has a dependency on the observer's distance from the black hole due to the non-flat asymptotic behaviour of the black hole spacetime. Hence, we investigated the shadow of the black hole with respect to the model parameters for different observer distances. The impacts of the model parameters $C$, $m$ and $c_1$ are more significant when the observer distance is large. The impacts of the other model parameters are almost independent of the observer distance. The impact of the noncommutative parameter $\theta$ on the black hole shadow is very small. We found that with an increase in the value of $\theta$, shadow radius $R_s$ decreases nonlinearly.

\section*{Acknowledgments}
P.B. is thankful to the Inter-University Centre for Astronomy and Astrophysics (IUCAA), Pune, Government of India, for providing visiting associateship. DJG acknowledges the contribution of the COST Action CA21136  -- ``Addressing observational tensions in cosmology with systematics and fundamental physics (CosmoVerse)". SP acknowledges funding support from the NSRF via the Program Management Unit for Human Resources \& Institutional Development, Research and Innovation [grant number B39G670016].
\section*{Data Availability Statement}
There are no new data associated with this article.

\bibliography{references}
\end{document}